\documentclass{elsarticle}
\usepackage{lineno,hyperref}
\usepackage[table]{xcolor}
\usepackage{amsmath}
\usepackage{amssymb}
\usepackage{latexsym}
\usepackage{framed,multirow}
\usepackage{booktabs}
\modulolinenumbers[1]
\usepackage{float}
\usepackage{natbib}

\usepackage{arydshln}

\begin{document}

\begin{frontmatter}


\title{United Adversarial Learning for Liver Tumor Segmentation and Detection of Multi-modality Non-contrast MRI}


\author[mymainaddress,mysecondaryaddress]{Jianfeng Zhao}

\author[mymainaddress]{Dengwang Li\corref{mycorrespondingauthor}}
\cortext[mycorrespondingauthor]{Corresponding author}
\ead{dengwang@sdnu.edu.cn}

\author[mysecondaryaddress]{Shuo Li\corref{mycorrespondingauthor}}
\ead{slishuo@gmail.com}
\address[mymainaddress]{Shandong Key Laboratory of Medical Physics and Image Processing, Shandong Institute of Industrial Technology for Health Sciences and Precision Medicine, School of Physics and Electronics, Shandong Normal University, Jinan, 250358, China}
\address[mysecondaryaddress]{Western University, London ON, Canada}

\begin{abstract}
Simultaneous segmentation and detection of liver tumors (hemangioma and hepatocellular carcinoma (HCC)) by using multi-modality non-contrast magnetic resonance imaging (NCMRI) are crucial for the clinical diagnosis. However, it is still a challenging task due to: (1) the HCC information on NCMRI is invisible or insufficient makes extraction of liver tumors feature difficult; (2) diverse imaging characteristics in multi-modality NCMRI causes feature fusion and selection difficult; (3) no specific information between hemangioma and HCC on NCMRI cause liver tumors detection difficult. In this study, we propose a united adversarial learning framework (UAL) for simultaneous liver tumors segmentation and detection using multi-modality NCMRI. The UAL first utilizes a multi-view aware encoder to extract multi-modality NCMRI information for liver tumor segmentation and detection. In this encoder, a novel {\itshape edge dissimilarity feature pyramid module} is designed to facilitate the complementary multi-modality feature extraction. Second, the newly designed {\itshape fusion and selection channel} is used to fuse the multi-modality feature and make the decision of the feature selection. Then, the proposed mechanism of {\itshape coordinate sharing with padding} integrates the multi-task of segmentation and detection so that it enables multi-task to perform united adversarial learning in one discriminator. Lastly, an innovative {\itshape multi-phase radiomics guided discriminator} exploits the clear and specific tumor information to improve the multi-task performance via the adversarial learning strategy. The UAL is validated in corresponding multi-modality NCMRI (i.e. T1FS pre-contrast MRI, T2FS MRI, and DWI) and three phases contrast-enhanced MRI of 255 clinical subjects. The experiments show that UAL gains high performance with the dice similarity coefficient of 83.63\%, the pixel accuracy of 97.75\%, the intersection-over-union of 81.30\%, the sensitivity of 92.13\%, the specificity of 93.75\%, and the detection accuracy of 92.94\%, which demonstrate that UAL has great potential in the clinical diagnosis of liver tumors.
\end{abstract}


\end{frontmatter}
\section{Introduction}

\begin{figure}
\includegraphics[width=1\textwidth]{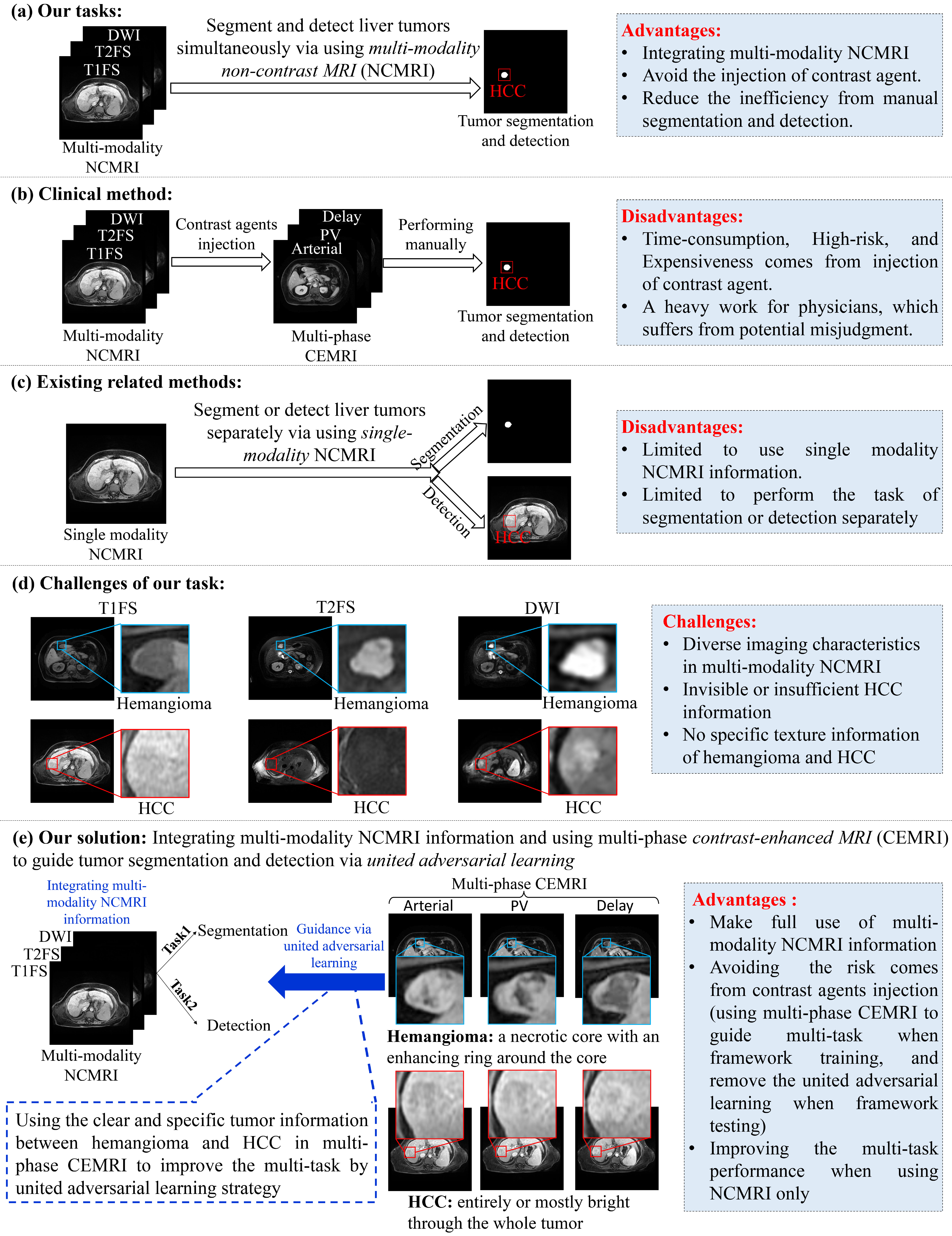}
\caption{Simultaneous liver tumors segmentation and detection via using multi-modality NCMRI are crucial for clinical diagnosis. However, existing related methods can not perform well due to some limitations. It is still challenging due to: 1) diverse imaging characteristics of multi-modality NCMRI cause multi-modality NCMRI fusion difficult; 2) invisible or insufficient HCC information in NCMRI makes feature extraction difficult; 3) no specific texture feature of hemangioma and HCC cause liver tumors detection difficult. To address these challenges, we proposed a united adversarial learning framework (UAL), which integrates multi-modality NCMRI information and utilizes the multi-phase CEMRI information to perform the adversarial learning for promoting the liver tumors segmentation and detection.} \label{fig1}
\end{figure}

Segmentation and detection of liver tumors (hemangioma and hepatocellular carcinoma (HCC)) of multi-modality non-contrast magnetic resonance imaging (NCMRI) is a time-saving, safe, and inexpensive solution for the clinical diagnosis and treatment \citep{kim2020diagnostic}, which as shown in Fig.\ref{fig1} (a). However, currently in clinical, the segmentation and detection of liver tumors would be performed manually by physicians via observing the multi-phase contrast-enhanced magnetic resonance imaging (CEMRI), which as shown in Fig.\ref{fig1} (b). It is heavy work that suffers from potential misjudgment. More importantly, the gadolinium contrast agents (CAs) injection of CEMRI suffers from time-consumption, high-risk and expensive \citep{idee2006clinical}. Especially for patients with compromised kidney function, they are restricted from injecting CAs\citep{marckmann2006nephrogenic}. Therefore, if the liver tumors segmentation and detection can be achieved via using NCMRI only, it will overcome the shortcomings in the clinic. Recently, there has been some related work trying to satisfy the clinical requirements, but they all have some limitations.

\subsection{Existing works for avoiding CAs injection}
Recently, to overcome the shortcomings caused by the injection of CAs, some works have attempted to segment or detect liver tumors via using single modality NCMRI, which as shown in Fig.\ref{fig1} (c). For instance, Xiao et al. \citep{xiao2019radiomics} attempted to segment liver tumors on T2FS via using the radiomics feature from delay-phase CEMRI. And Zhao et al. \citep{zhao2020tripartite} attempted to detect liver tumors on synthetic delay-phase CEMRI by using a Tripartite-GAN. However, the assumption of their works ignored the situation of HCC information is invisible or insufficient on T1FS and T2FS. So that the performance of their works will be limited when HCC is small or invisible due to lacking the information of DWI. Moreover, clinical research has shown some HCC cases that more clear on arterial-phase CEMRI than delay-phase CEMRI \citep{cereser2010comparison}. And the multi-phase CEMRI (i.e. arterial-phase, portal-venous (PV) phase, and delay-phase) has shown high sensitivity and specificity for clinical liver tumors diagnosis \citep{yu1999contrast, kierans2016diagnostic}. Therefore, the performance of \citep{xiao2019radiomics} and \citep{zhao2020tripartite} will be limited because they are all using delay-phase CEMRI only. Lastly, it is crucial for clinical that segment liver tumors and detect the tumor whether is benign or malignant simultaneously. But their works are limited to segment or detect liver tumors separately.

\subsection{Challenges}
For liver tumors segmentation and detection, the segmentation task pays more attention to the accurate boundary information while the detection task focuses more on the location, size, and specific information of liver tumors. Therefore, simultaneously segment and detect liver tumors by using multi-modality NCMRI (i.e. T1FS, T2FS, and DWI) is still challenging due to some limitations, which as shown in Fig.\ref{fig1}(d). 1) Diverse imaging characteristics in multi-modality NCMRI causes the feature fusion and section difficult for multi-task. 2) HCC information in NCMRI is invisible or insufficient \citep{choi2014ct} limits the performance of segmentation and detection. 3) No specific texture information of liver tumors has the risk of confusing liver tumors.

\subsection{Our proposed method}
In this study, we propose a united adversarial learning framework (UAL) to simultaneously segment and detect liver tumors by using multi-modality NCMRI, which is shown in Fig.\ref{fig1} (d). Our basic assumption is that the integration of complementary information from multi-modality MRI can enhance the feature representation and using multi-phase radiomics (MPR) features from CEMRI to perform adversarial learning can guide the detailed information extraction of NCMRI. And then it will improve the performance of liver tumors segmentation and detection. Specifically, the UAL first utilizes three parallel convolution channels for multi-modality NCMRI information extraction. To facilitate the complementary multi-modality feature extraction, a novel edge dissimilarity feature pyramid module (EDFPM) is designed to extract the multi-size edge dissimilarity maps. Additionally,  the multi-size edge dissimilarity maps as the prior knowledge added into the convolution channel make the UAL easy to training. Then, a fusion and selection channel (FSC) is designed to make the final decision of feature fusion and selection. After FSC, we proposed the mechanism of coordinate sharing with padding (CSWP) to integrate the segmentation task and detection task so that it enables multi-task to perform united adversarial learning. Lastly, an innovative multi-phase radiomics guided discriminator (MPRG-D) exploits the clear and specific tumor information to improve the multi-task performance via the adversarial learning strategy.

The contributions of this work are summarized as following:
\begin{itemize}
\item For the first time, the proposed UAL provided a time-saving, safe, and inexpensive tool, which achieves simultaneous liver tumors segmentation and detection via using multi-modality NCMRI only, especially for HCC information is invisible or insufficient.
\item The novel EDFPM extracts the multi-size edge dissimilarity maps to enhance the multi-modality NCMRI feature extraction and it as the prior knowledge added into UAL facilitates the UAL training.
\item The innovative FSC fuses the multi-modality NCMRI feature and adaptively makes the final decision of feature selection according to the feature requirements of liver tumors segmentation task and detection task.
\item We proposed a CSWP mechanism, which enables the MPRG-D achieves the united adversarial learning for liver tumors segmentation and detection for the first time.
\item The newly designed MPRG-D enhances discrimination by adding the MPR feature. In this situation, it is capable of constraint accurate liver tumors segmentation and detection via the adversarial learning strategy.
\end{itemize}

\section{Related work}
\subsection{Deep learning for automatic liver tumors diagnosis}
Deep learning for automatic liver tumors diagnosis has attracted great interests since it can extract high-level semantic information \citep{bousabarah2020automated}. Hamm et al. \citep{hamm2019deep} developed and validated a convolutional neural network-based deep learning system that classifies common liver lesions on CEMRI. Bousabarah et al. \citep{bousabarah2020automated} adopted a deep convolutional neural network with the radiomics feature to automatically detect and delineate HCC on CEMRI. Kim et al. \citep{kim2020detection} used a deep learning-based classifier to detect HCC on CEMRI. All these clinical researches demonstrated that deep learning can provide an effective solution for liver tumors diagnosis. However, all of these models are trained and tested based on CEMRI, which suffers from CAs injection.

Recently, to overcome the shortcomings caused by the injection of CAs. For instance, the work of Xiao et al. \citep{xiao2019radiomics} attempted to extract T2FS information and perform adversarial learning via using delay-phase CEMRI information for liver tumors segmentation. The work of Zhao et al. \citep{zhao2020tripartite} attempted to learn the highly non-linear mapping between T1FS and CEMRI for liver tumors detection. However, their works do not perform well due to using single modality NCMRI and single CEMRI. Besides, both of their works are limited to segment or detect liver tumors separately.

\subsection{Multi-task of simultaneous segmentation and detection}
Some deep learning-based networks used for multi-task of simultaneous segmentation and detection have achieved great success \citep{wu2007simultaneous, hariharan2014simultaneous, he2017mask, gao2020feature}. The work \citep{he2017mask} proposed a general framework named Mask R-CNN for object instance segmentation. The work \citep{gao2020feature} proposed a derivative of Mask R-CNN for breast cancer segmentation and detection. However, they are not performed-well when simultaneous segmentation and detection of small lesions. Because the feature maps used for segmentation are highly reduced in spatial resolution \citep{gao2020feature}. 
\subsection{Clinical researches of liver tumors diagnosis}
Nowadays, clinical researches on liver tumors diagnosis showed that the complementary information between multi-modality NCMRI has high sensitivity when used for HCC diagnosis \citep{han2018diagnostic, wu2019radiomics, canellas2019lesion}. DWI with clear location information of liver lesions has shown excellent performance in detecting liver tumors \citep{kele2010diffusion, piana2011new, kim2012hypovascular, vandecaveye2009diffusion}. The research of \citep{ebeed2017role, xu2009added} proved that DWI with multi-modality CEMRI helped to provide higher sensitivities than multi-modality CEMRI alone in the detection of HCC.

\textbf{\itshape Discussions:} To the best of our knowledge, no work attempted to segment and detect liver tumors simultaneously via using multi-modality NCMRI. Therefore, if liver tumors segmentation and detection can be completed via using multi-modality NCMRI only. It will greatly optimize the liver tumors diagnosis and overcome the shortcomings of existing works.

\section{Methodology}
\begin{figure*}[h]
\includegraphics[width=1\textwidth]{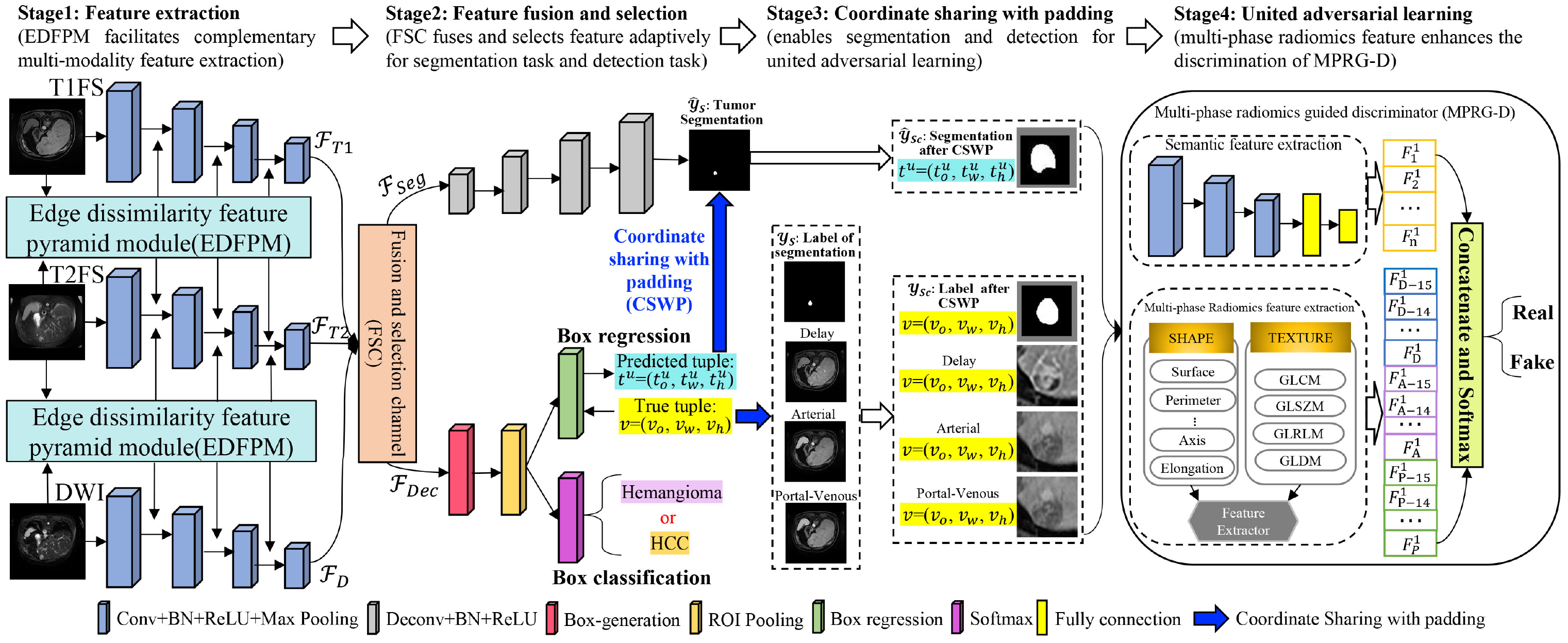}
\caption{Overview of the proposed UAL. It segments and detects liver tumors simultaneously contains four stages. 1) Using three parallel convolution channels to extract the multi-modality NCMRI feature with the help of newly designed EDFPM. (2) To solve the challenge of multi-modality fusion, UAL using an innovative FSC to fuse and select features according to the feature requirements of the segmentation task and detection task. 3) To perform the united learning strategy for promoting liver tumors segmentation and detection, UAL using the CSWP mechanism to unify the outputs of segmentation and detection. 4) To solve the challenge of HCC is invisible and no specific texture feature of hemangioma and HCC, we proposed an MPRG-D which is added the multi-phase radiomics from CEMRI to perform the united adversarial learning strategy.} \label{UAL}
\end{figure*}

The UAL integrates multi-modality NCMRI information for liver tumors segmentation and detection. Moreover, it uses multi-phase CEMRI to guide segmentation and detection by adversarial learning strategy when framework training. And it avoids the risk that comes from contrast agents injection by removing the adversarial learning when framework testing. Specifically, as shown in Fig.\ref{UAL}, the UAL is fed with T1FS ($\mathcal{X}^{T1}\in \mathbb{R}^{H\times W\times N}$), T2FS ($\mathcal{X}^{T2}\in \mathbb{R}^{H\times W\times N}$), and DWI ($\mathcal{X}^{D}\in \mathbb{R}^{H\times W\times C}$) sequences, and outputs the results of liver tumor segmentation ${\hat{\mathcal{Y}}}_S$ and detection \{$\mathcal{Y}_p, t^u$\}. The UAL is performed via the following four stages:
\begin{itemize}
\item An encoder using three parallel convolution channels with {\itshape\textbf{EDFPM}} for multi-modality NCMRI feature extraction. The EDFPM extracts the multi-size edge dissimilarity maps, which as the prior knowledge added into the three parallel convolution channels to facilitate complementary multi-modality NCMRI information extraction.
\item The {\itshape\textbf{FSC}} fuses the feature from the three parallel convolution channels and makes the final decision of feature selection for liver tumors segmentation and detection.
\item The {\itshape\textbf{CSWP}} integrates the outputs of liver tumors segmentation and detection via the operation of padding using number 2. And then feeding the integration into discriminator, which enables the united adversarial learning for multi-task. 
\item The {\itshape\textbf{MPRG-D}} extracts the semantic feature and MPR feature to distinguish the real or fake of the integration. And then constrain the multi-task to optimize by adversarial learning strategy.
\end{itemize}

\subsection{EDFPM for multi-modality NCMRI feature extraction.}
\begin{figure}[h]
\includegraphics[width=1\textwidth]{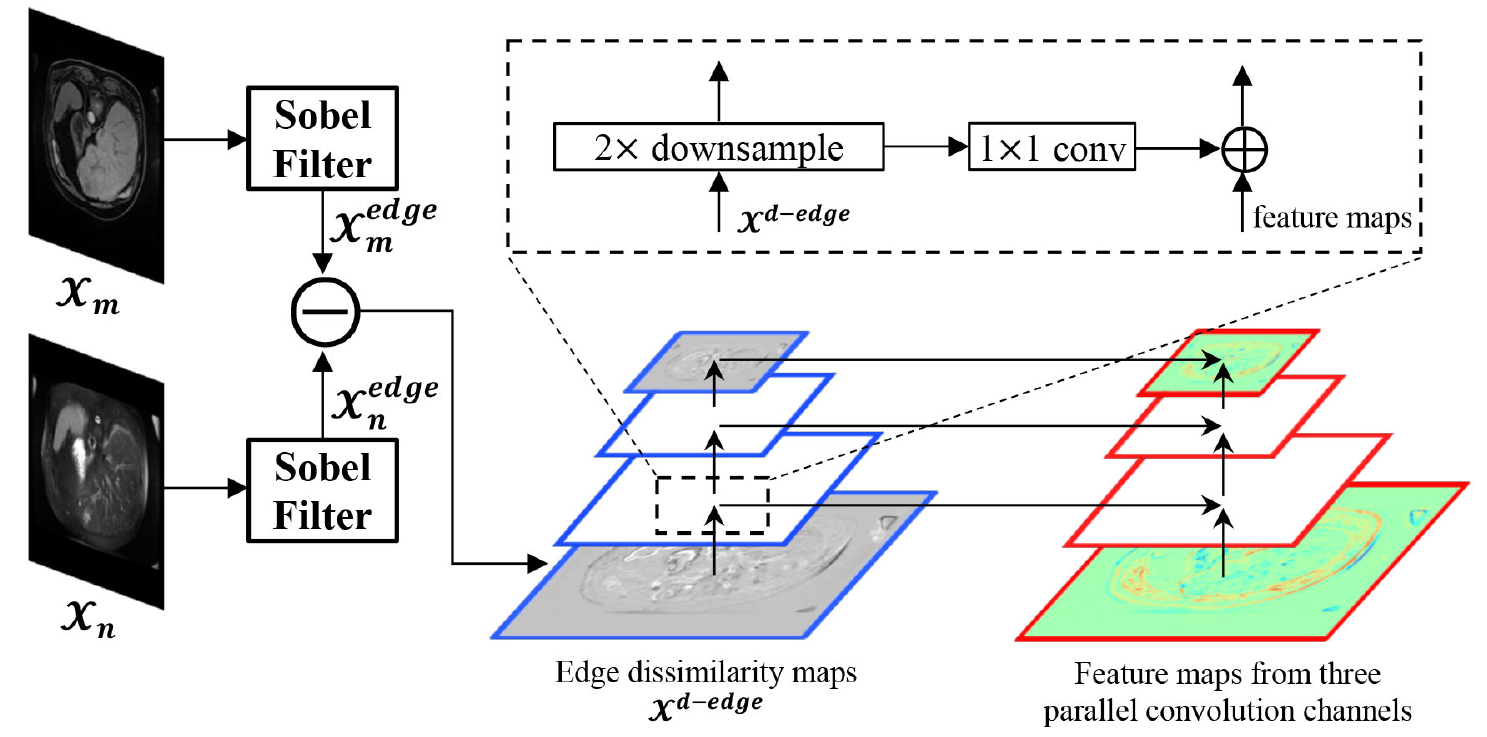}
\caption{The calculation mechanism of EDFPM. It yields multi-size edge dissimilarity maps via using Sobel Filter and downsample operation, in which the pyramid structure refines the spatial precision of the edge dissimilarity maps. And adding the multi-size edge dissimilarity maps into UAL to promote multi-modality NCMRI feature extraction.} \label{fig3}
\end{figure}

As shown in Fig.\ref{fig3},  the calculation mechanism of EDFPM mainly goes through three steps. (1) Feeding two modalities MRI of $\mathcal{X}_m, \mathcal{X}_n\in\{\mathcal{X}^{T1}, \mathcal{X}^{T2}, \mathcal{X}^{D}\}$ into the Sobel-based edge detector \citep{sobel19683x3} to generate the corresponding edge maps $\mathcal{X}_m^{edge}$ and $\mathcal{X}_n^{edge}$. Then perform the element-wise subtraction between $\mathcal{X}_m^{edge}$ and $\mathcal{X}_n^{edge}$ to yield the edge dissimilarity maps $\mathcal{X}^{d-edge}$. (2) The bilinear interpolation is used to perform the operation of $2\times{downsample}$, which ensures the multi-size edge dissimilarity maps are consistent with the size of the feature maps from the connected convolution layer. Besides, the pyramid structure refines the spatial precision of the edge dissimilarity maps. (3) The connection manner of multi-size edge dissimilarity maps and feature maps is shown in the dashed window of Fig.\ref{fig3}. The 1$\times$1 convolution operation is added to the path of the skip connection. It is used to change the channel number of edge dissimilarity maps to make it consistent with that of feature maps from three parallel convolution channels. The edge dissimilarity maps promote the complementary multi-modality feature extraction while also accelerating the convergence of the UAL.

\subsection{FSC for multi-modality feature fusion and selection.}
\begin{figure}[h]
\includegraphics[width=1\textwidth]{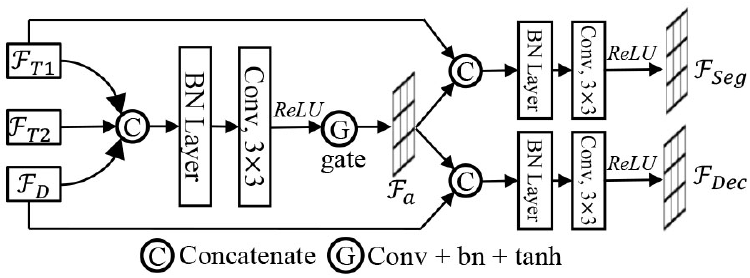}
\caption{The calculation mechanism of FSC. It mainly contains a series of operations for multi-modality fusion, a gate for feature selection, and the weight setting of the multi-modality NCMRI.} \label{FSC}
\end{figure}
As the stage2 shown in Fig.\ref{UAL}, the FSC is connected at the end of $Encoder$ for feature $\mathcal{F}^{T1}$, $\mathcal{F}^{T2}$, and $\mathcal{F}^{D}$ fusion and selection. The calculation mechanism of FSC is shown in Fig.\ref{FSC}. The FSC composes a gate for controlling the information fusion and selection, in which the gate has been proposed by \citep{pang2019direct} and developed in \citep{ge2019k}. In this work, we are the first to design FSC with the gate for three modalities MRI fusion. Moreover, according to the feature requirements of multi-task, FSC increases the weight of the dominant modality MRI. In this way, it can adaptively integrate and select features to improve multi-task performance. Specifically, it mainly goes on two steps. (1) Obtaining the preliminary fusion features $\mathcal{F}^{a}$ of the $\mathcal{F}^{T1}$, $\mathcal{F}^{T2}$, and $\mathcal{F}^{D}$ without setting any weight. It can be calculated as:

\begin{equation}
\mathcal{F}_a=\sigma(\varepsilon(x^{[\mathcal{F}^{T1}, \mathcal{F}^{T2}, \mathcal{F}^{D}]}*\mathcal{W}_a+b_a)*\mathcal{W}_b+b_b)
\end{equation}
where the features of $x^{[\mathcal{F}^{T1}, \mathcal{F}^{T2}, \mathcal{F}^{D}]}$ represents the concatenation of $\mathcal{F}^{T1}, \mathcal{F}^{T2}$, and $\mathcal{F}^{D}$. $\mathcal{W}_a$, $b_a$, $\mathcal{W}_b$ and $b_b$ are the trainable weights and biases of the first two convolution operations, $\varepsilon$ is $ReLU$ activation, $\sigma$ is $tanh$ activation. (2) Obtaining the final fusion features $\mathcal{F}_{Seg}$ for segmentation path and $\mathcal{F}_{Dec}$ for detection path. Inspired by \citep{leng2018context} that context information will improve the segmentation performance, we concatenate the $\mathcal{F}^{a}$ and $\mathcal{F}^{T1}$ to increases the weight of global anatomical information on T1FS when calculating $\mathcal{F}_{Dec}$. And inspired by \citep{kele2010diffusion, piana2011new, kim2012hypovascular, vandecaveye2009diffusion} that DWI with clear location information of liver lesions has shown excellent performance in detecting liver tumors, we concatenate the $\mathcal{F}^{a}$ and $\mathcal{F}^{D}$ to increases the weight of location information on DWI when calculating $\mathcal{F}_{Dec}$. The calculation of $\mathcal{F}_{Seg}$ and $\mathcal{F}_{Dec}$ as follows:

\begin{equation}
\mathcal{F}_{Seg}=\varepsilon(x^{[\mathcal{F}_a, \mathcal{F}^{T1}]}*\mathcal{W}_c+b_c)
\end{equation}
\begin{equation}
\mathcal{F}_{Dec}=\varepsilon(x^{[\mathcal{F}_a, \mathcal{F}^{D}]}*\mathcal{W}_d+b_d)
\end{equation}
where the features of $x^{[\mathcal{F}_a, \mathcal{F}^{T1}]}$ represents the concatenation of $\mathcal{F}^{T1}$ and feature maps $\mathcal{F}_a$, $x^{[\mathcal{F}_a, \mathcal{F}^{D}]}$ represents the concatenation of $\mathcal{F}^{D}$ and feature maps $\mathcal{F}_a$, $\mathcal{W}_c$ and $b_c$ are the trainable weights and biases of the last convolution operation for segmentation path, and $\mathcal{W}_d$ and $b_d$ are the trainable weights and biases of the last convolution operation for detection path.

\subsection{CSWP integrates liver tumors segmentation and detection for united adversarial learning.}
\begin{figure}[h]
\includegraphics[width=1\textwidth]{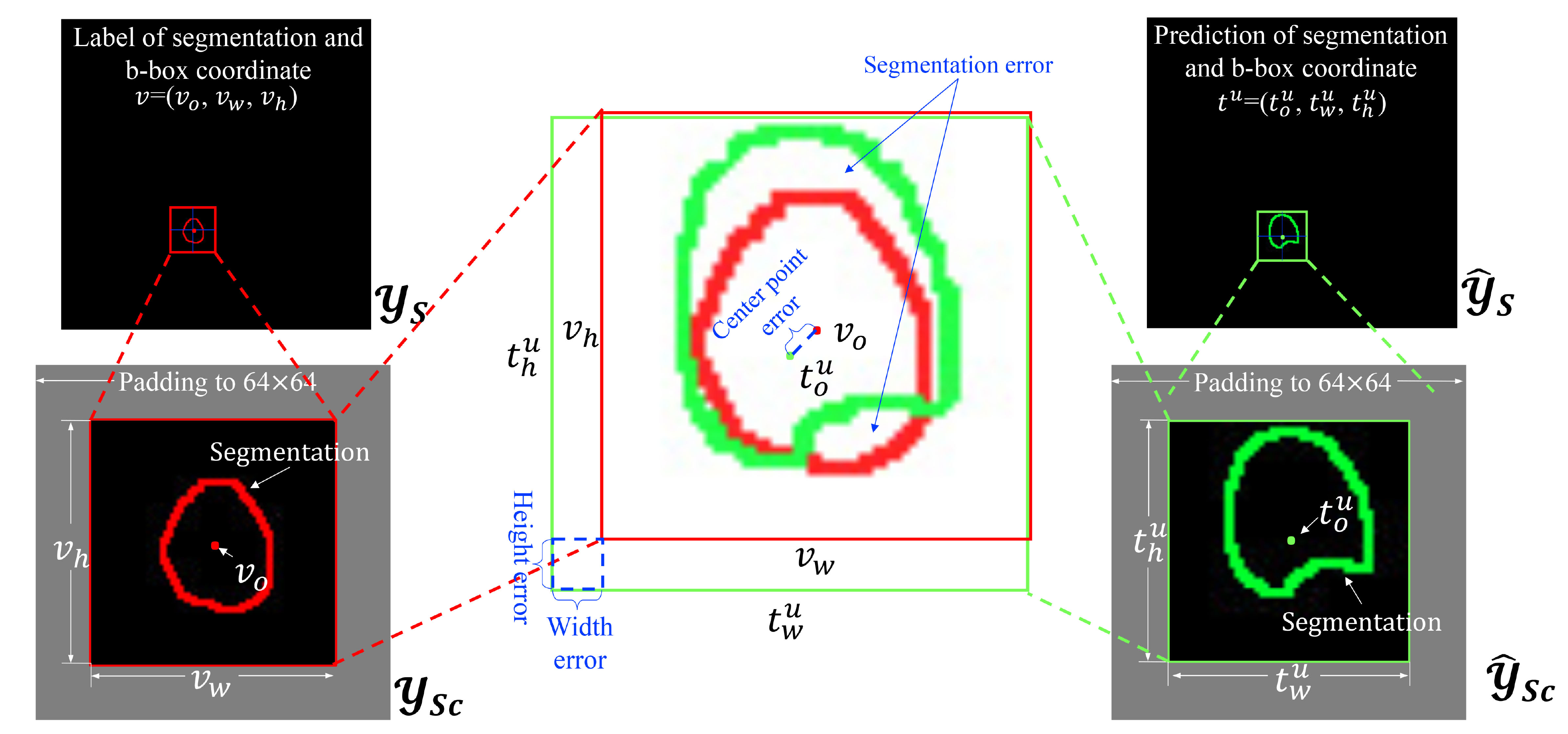}
\caption{The mechanism of CSWP. It uses coordinate sharing to unify the outputs of segmentation task and detection task. And in the premise of fixed coordinates, use the padding operation to unify their size for feature extraction.} \label{CSWP}
\end{figure}

The CSWP enables a united adversarial learning for multi-task of segmentation and detection for the first time. The calculation mechanism of CSWP as shown in Fig.\ref{CSWP}. First, we defined the label of segmentation image is $\mathcal{Y}_S$, the prediction of segmentation image is ${\hat{\mathcal{Y}}}_S$, the true tuple as the ground-truth of the bounding-box (b-box) is $v=(v_o, v_w, v_h)$, and the predicted tuple of b-box regression is $t_u=(t_o^u, t_w^u, t_h^u)$. Where the $v_o$ and $t_o^u$ are the center point of b-box, $v_w=v_h$ means the true b-box is square, $t_w^u=max\{t_w^u, t_h^u\}$ and $t_h^u=max\{t_w^u, t_h^u\}$ makes the the predicted b-box is square. Then, to integrate the $\mathcal{Y}_S$ and $v$ for feature extraction in MPRG-D, using the operation of number 2 padding and taking the $v_o$ as the center for generating the integration $\mathcal{Y}_{Sc}$. Performing the same operation on ${\hat{\mathcal{Y}}}_S$ and $t_u$ to obtain the integration ${\hat{\mathcal{Y}}}_{Sc}$. The padding operation makes $\mathcal{Y}_{Sc}$ and ${\hat{\mathcal{Y}}}_{Sc}$ to $64\times64$ pixels. In this situation, CSWP enables segmentation and detection can be unified to be optimized by performing adversarial learning in our MPRG-D with $\mathcal{L}_{adv}({\hat{\mathcal{Y}}}_{Sc}, \mathcal{Y}_{Sc})$. And then reduce the error (i.e. segmentation error, center point error, height error, and width error) between ${\hat{\mathcal{Y}}}_{Sc}$ and $\mathcal{Y}_{Sc}$ via the united adversarial learning strategy.

\subsection{MPRG-D for united adversarial learning.}
As the stage4 shown in Fig.\ref{UAL}, the MPRG-D receives ${\hat{\mathcal{Y}}}_{Sc}$ and $\mathcal{Y}_{Sc}$ after the operation of CSWP and outputs a single scalar that the real or fake of ${\hat{\mathcal{Y}}}_{Sc}$ and $\mathcal{Y}_{Sc}$. Then the discrimination of MPRG-D is fed back to the segmentation path and detection path via the adversarial strategy with $\mathcal{L}_{adv}({\hat{\mathcal{Y}}}_{Sc}, \mathcal{Y}_{Sc})$. To enhance the discrimination and fully extract features of liver tumors, MPR from multi-phase CEMRI is added into MPRG-D. Because the multi-phase CEMRI has shown high sensitivity and specificity for clinical liver tumors diagnosis \citep{yu1999contrast, kierans2016diagnostic}. Specifically, the MPRG-D utilizes a network with three convolutional layers to extract semantic features and utilizes a python toolbox Pyradiomics \citep{van2017computational} to extract multi-phase radiomics feature. For semantic feature extraction, the input is ${\hat{\mathcal{Y}}}_{Sc}$ or $\mathcal{Y}_{Sc}$. And for MPR feature extraction, the input is ${\hat{\mathcal{Y}}}_{Sc}$ with multi-phase CEMRI or $\mathcal{Y}_{Sc}$ with multi-phase CEMRI. Finally, the output of real of fake is obtained by feeding the concatenation of the semantic feature and MPR feature to $Softmax$ layer. With learn through this adversarial strategy, the MPRG-D can constrain our UAL to train an excellent model with accurate liver tumors segmentation and detection.

\subsection{Constraint strategy of UAL.}
The UAL is the first time to achieves simultaneously promote the segmentation and detection via using a united adversarial learning strategy. The basic adversarial learning strategy of our UAL is coming from the minimax game between generator and discriminator in the primary GAN \citep{goodfellow2014generative}, which the minimax optimization formulated as: 

\begin{equation}
{\min \limits_{G}} \max \limits_{D} [E_{y\sim p_{data}(y)}[(logD(y)]+E_{x\sim p_{x}(x)}[log(1-D(G(x))]]
\end{equation}
UAL is trained to minimize the probability of ${\hat{\mathcal{Y}}}_{Sc}$ to be recognized while maximizing the probability of making mistakes of the discriminator when discriminating the ${\hat{\mathcal{Y}}}_{Sc}$ It means that the $y\sim p_{data}(y)$ in equation(4) corresponding to $\mathcal{Y}_{Sc}$ and $G(x)$ equation(4) corresponding to ${\hat{\mathcal{Y}}}_{Sc}$. Specifically, for the optimization of tumors segmentation, the loss function $\mathcal{L}_{Seg}$ is defined as:

\begin{equation}
\mathcal{L}_{Seg}({\hat{\mathcal{Y}}}_S, \mathcal{Y}_S, {\hat{\mathcal{Y}}}_{Sc})=\mathcal{L}_{pix-CE}({\hat{\mathcal{Y}}}_S, \mathcal{Y}_S)+\lambda_1\mathcal{L}_{adv}(\mathcal{D}({\hat{\mathcal{Y}}}_{Sc}, 1))
\end{equation}
where the hyper-parameter $\lambda_1$ set to one for maintaining the balance between tumor segmentation $\mathcal{L}_{pix-CE}$ and the adversarial learning $\mathcal{L}_{adv}$. In which $\mathcal{L}_{pix-CE}$ and $\mathcal{L}_{adv}$ are:

\begin{equation}
\begin{aligned}
&\mathcal{L}_{pix-CE}({\hat{\mathcal{Y}}}_S, \mathcal{Y}_S) = \\ 
&\frac{\sum_{n}\sum_{i, j}({\hat{\mathcal{Y}}}^{i,j}_n log(\mathcal{Y}^{i,j}_n)+(1-{\hat{\mathcal{Y}}}^{i,j}_n) log(1-\mathcal{Y}^{i,j}_n))}{NHW}
\end{aligned}
\end{equation}
\begin{equation}
\mathcal{L}_{adv}({\hat{\mathcal{Y}}}_{Sc}, \mathcal{Y}_{Sc})=-\sum_{n}\mathcal{Y}_{Sc}^nlog({\hat{\mathcal{Y}}}_{Sc}^n)+(1-\mathcal{Y}_{Sc}^n)log(1-{\hat{\mathcal{Y}}}_{Sc}^n)
\end{equation}
where ${\hat{\mathcal{Y}}}^{i,j}_n$ and $\mathcal{Y}^{i,j}_n$ represents the pixel classification located at $(i, j)$ in ${\hat{\mathcal{Y}}}_S$ and $\mathcal{Y}_S$. $n\in$\{$\mathcal{Y}_S$\} represents the each image of $\mathcal{Y}^{i,j}_S$. $H$ and $W$ is the height and the width of each image $n$ from the total training images of $N$.

The loss function of $\mathcal{L}_{D}$ for MPRG-D optimization is defined as:
\begin{equation}
\mathcal{L}_{D}({\hat{\mathcal{Y}}}_{Sc}, \mathcal{Y}_{Sc})=\mathcal{L}_{adv}(\mathcal{D}({\hat{\mathcal{Y}}}_{Sc}), 1)+\mathcal{L}_{adv}(\mathcal{D}(\mathcal{Y}_{Sc}), 0)
\end{equation}

The loss function of $\mathcal{L}_{Dec}$ for liver tumors detection optimization is defined as:

\begin{equation}
\begin{aligned}
\mathcal{L}_{Dec}(\mathcal{Y}_p, \mathcal{Y}_u, t^u, v, {\hat{\mathcal{Y}}}_{Sc})=&\mathcal{L}_{cls}(\mathcal{Y}_p, \mathcal{Y}_u)+\lambda_2[u\ge1]\mathcal{L}_{reg}(t^u, v)\\
&+\lambda_3\mathcal{L}_{adv}(\mathcal{D}({\hat{\mathcal{Y}}}_{Sc}, 1))
\end{aligned}
\end{equation}
where the hyper-parameter $\lambda_2$ and $\lambda_3$ set to one for maintaining the balance of the adversarial loss $\mathcal{L}_{adv}$ and two tasks losses of tumor classification $\mathcal{L}_{cls}$ and b-box regression $\mathcal{L}_{reg}$. The $\mathcal{L}_{cls}$ and $\mathcal{L}_{reg}$ are:

\begin{equation}
\left\{
\begin{aligned}
&\mathcal{L}_{cls}(\mathcal{Y}_p, \mathcal{Y}_u)=-logp_{\mathcal{Y}_u} \\ 
&\mathcal{L}_{reg}(t^u, v)=\sum_{i\in{\{o,w,h\}}}smooth_{L_1}(t_i^u-v_i)
\end{aligned}
\right.
\end{equation}
in which $smooth_{\mathcal{L}_1}(x)=0.5x^2$ if $|x|\textless{1}$, and $smooth_{\mathcal{L}_1}(x)=|x|-0.5$ otherwise. Where the $\mathcal{Y}_p$ represents the outputs of the probability distribution of liver tumors, $\mathcal{Y}_u$ represents the ground-truth class (i.e. hemangioma or HCC), the $[\mathcal{Y}_u\ge1]$ evaluates to 1 when $\mathcal{Y}_u\ge1$ and 0 otherwise.\\

We show the architecture details of our UAL in Table.\ref{Implementation}. And the \textbf{Algorithm 1} summarizes the procedure of UAL. The algorithm contains two parts: the training process of the UAL and the testing process of UAL. In which the training process of UAL mainly contains four stages for forward propagation and backward propagation for parameters update. The testing process of UAL mainly goes through two stages for liver tumors segmentation and detection.

\begin{table}[h]
\caption{The architecture of the UAL. The "Params" include: (1) Kernel size; (2) "@": Number of channels; (3) "Pad": Spatial padding number; (4) "Str": stride number. (Conv means convolution, Deconv means deconvolution, FC means fully connection)}\label{Implementation}
\begin{tabular}{|c|c|c|}
\hline
Stage of UAL                                                                                                   & \begin{tabular}[c]{@{}c@{}}Layer\\ Annotation\end{tabular} & Params                  \\ \hline
\multirow{4}{*}{\begin{tabular}[c]{@{}c@{}}Feature extraction\\ in Stage1 \\(i$\in$\{T1FS, T2FS, DWI\})\end{tabular}} & $Conv_1^{i}$                                                      & 3*3@64; Pad: 1; Str: 1  \\ \cline{2-3} 
                                                                                                        & $Conv_2^{i}$                                                    & 3*3@128; Pad: 1; Str: 1 \\ \cline{2-3} 
                                                                                                        & $Conv_3^{i}$                                                    & 3*3@256; Pad: 1; Str: 1 \\ \cline{2-3} 
                                                                                                        & $Conv_4^{i}$                                                     & 3*3@512; Pad: 1; Str: 1 \\ \hline
\multirow{4}{*}{\begin{tabular}[c]{@{}c@{}}Decoder of $\mathcal{F}_{Seg}$\\ in Stage2 \end{tabular}}                                                      & $Decond1$                                                    & 3*3@512; Pad: 1; Str: 1 \\ \cline{2-3} 
                                                                                                        & $Decond2$                                                    & 3*3@256; Pad: 1; Str: 1 \\ \cline{2-3} 
                                                                                                        & $Decond3$                                                    & 3*3@128; Pad: 1; Str: 1 \\ \cline{2-3} 
                                                                                                        & $Decond4$                                                    & 3*3@64; Pad: 1; Str: 1  \\ \hline
\multirow{5}{*}{\begin{tabular}[c]{@{}c@{}}Semantic feature extraction\\ in Stage4 \end{tabular}}                                                               & $Conv5$                                                      & 3*3@64; Pad: 1; Str: 1  \\ \cline{2-3} 
                                                                                                        & $Conv6$                                                      & 3*3@128; Pad: 1; Str: 1 \\ \cline{2-3} 
                                                                                                        & $Conv7$                                                      & 3*3@256; Pad: 1; Str: 1 \\ \cline{2-3} 
                                                                                                        & $FC1$                                                        & 1*1@256                 \\ \cline{2-3} 
                                                                                                        & $FC2$                                                        & 1*1@1                   \\ \hline
\end{tabular}
\end{table}

\begin{table}[]
\renewcommand\arraystretch{1.13}
\begin{tabular}{ l } 
\toprule
\textbf{Algorithm 1:} \\
\midrule
\textbf{Training process of UAL:}\\
\ \ \textbf{Input:} Multi-modality NCMRI $\mathcal{X}^{T1}$, $\mathcal{X}^{T2}$, and $\mathcal{X}^D$; Arterial-phase \\
\ \ CEMRI $\mathcal{X}^{A}$; PV-phase CEMRI $\mathcal{X}^{P}$; Delay-phase CEMRI $\mathcal{X}^{De}$; \\
\ \ Segmentation label $\mathcal{Y}_S$; True tuple of b-box $v$; The label of the types \\
\ \ of tumors $\mathcal{Y}_u$; Loss balanced weights $\lambda_1$, $\lambda_2$, and $\lambda_3$; Batchsize n;\\
\ \ Learning rate $\eta$; Iteration number M.\\
\ \ \textbf{Output:} Learned parameters \{$\theta_{seg}, \theta_{Dec}, \theta_{Dis}$\} \\
\ \ \textbf{for} step in \textbf{M} do\\
\ \ \ \ \textbf{begin} forward propagation:\\
\ \ \ \ \ \ \textbf{Stage1:} Feature extraction\\
\ \ \ \ \ \ $\{\mathcal{F}^{T1}, \mathcal{F}^{T2}, \mathcal{F}^{D}\}=Encoder(\mathcal{X}^{T1}_n$, $\mathcal{X}^{T2}_n$, $\mathcal{X}^D_n) $\\
\ \ \ \ \ \ \textbf{Stage2:} Feature fusion and selection for segmentation and detection\\
\ \ \ \ \ \ $\{\mathcal{F}^{Seg}, \mathcal{F}^{Dec}\}=FSC(\mathcal{F}^{T1}, \mathcal{F}^{T2}, \mathcal{F}^{D})$\\
\ \ \ \ \ \ ${\hat{\mathcal{Y}}}_S=SegPath(\mathcal{F}^{Seg})$\\
\ \ \ \ \ \ $\{\mathcal{Y}_p, t^u\}=DecPath(\mathcal{F}^{Dec})$\\
\ \ \ \ \ \ \textbf{Stage3:} Coordinate sharing with padding\\
\ \ \ \ \ \ $\hat{\mathcal{Y}}_{Sc}=CSWP({\hat{\mathcal{Y}}}_S, t^u)$ \\
\ \ \ \ \ \ $\mathcal{Y}_{Sc}=CSWP(\mathcal{Y}_S, v)$ \\
\ \ \ \ \ \ \textbf{Stage4:} United adversarial learning\\
\ \ \ \ \ \ $D(\hat{\mathcal{Y}}_{Sc})=$$MPRG$-$D(\hat{\mathcal{Y}}_{Sc})$\\
\ \ \ \ \ \ $D(\mathcal{Y}_{Sc})=$$MPRG$-$D(\mathcal{Y}_{Sc})$\\
\ \ \ \ \textbf{end}\\
\ \ \ \ \textbf{begin} backward propagation:\\
\ \ \ \ \ \ $\theta_{seg}=\theta_{seg}-\eta \nabla(\mathcal{L}_{pix-CE}({\hat{\mathcal{Y}}}_S, \mathcal{Y}_S)+\lambda_1\mathcal{L}_{adv}(\mathcal{D}({\hat{\mathcal{Y}}}_{Sc}, 1)))$\\
\ \ \ \ \ \ $\theta_{Dec}= \theta_{Dec}-\eta \nabla(\mathcal{L}_{cls}(\mathcal{Y}_p, \mathcal{Y}_u)+\lambda_2[u\ge1]\mathcal{L}_{reg}(t^u, v)$\\
\ \ \ \ \ \ \ \ \ \ \ \ \ \ \ \ \ \ \ \ \ \ \ \ \ \ \ \ \ \ \ \ \ \ \ \ \ \ \ \ \ \ \ \ \ \ \ \ \ $+\lambda_3\mathcal{L}_{adv}(\mathcal{D}({\hat{\mathcal{Y}}}_{Sc}, 1)))$\\
\ \ \ \ \ \ $\theta_{Dis}= \theta_{Dis}-\eta \nabla(\mathcal{L}_{D}(\mathcal{L}_{adv}(\mathcal{D}({\hat{\mathcal{Y}}}_{Sc}), 1)+\mathcal{L}_{adv}(\mathcal{D}(\mathcal{Y}_{Sc}), 0))$\\
\ \ \ \ \textbf{end}\\
\ \ \textbf{end}\\
\textbf{Testing process of UAL:}\\
\ \ \textbf{Stage1:} fed multi-modality NCMRI $\mathcal{X}^{T1}$, $\mathcal{X}^{T2}$, and $\mathcal{X}^D$\\
\ \ \textbf{Stage2: Prediction of liver tumors segmentation and detection}\\
\ \ forward propagate the $\mathcal{X}^{T1}$, $\mathcal{X}^{T2}$, and $\mathcal{X}^D$ through UAL with trained\\
\ \ weights, and get the prediction of liver tumors segmentation $\hat{\mathcal{Y}}_S$ and \\
\ \ detection $\{\mathcal{Y}_p, t^u\}$. \\
\bottomrule
\end{tabular}
\end{table}

\section{Experiments}
The effectiveness of the proposed UAL is validated in the liver tumors segmentation and detection. Experimental results show that UAL successfully segments and detects liver tumors via using multi-modality NCMRI, and achieves dice similarity coefficient (DSC) of 83.63\%, pixel accuracy (p-Acc) of 97.75\%, intersection-over-union (IoU) of 81.30\%, the sensitivity of 92.13\%, the specificity of 93.75\%, and detection accuracy of 92.94\%.

\subsection{Dataset and Configurations} Our UAL is validated on a clinical dataset with totaling 255 subjects (125 subjects of hemangioma and 130 subjects of HCC), and each subject has corresponding T1FS [256$\times$256 px], T2FS [256$\times$256 px], DWI [256$\times$256 px] and multi-phase CEMRI [256$\times$256 px] collected after standard clinical liver MRI examinations. CEMRI used in these protocols was gadobutrol 0.1 mmol/kg on a 3T MRI scanner. The segmentation labels performed on CEMRI are obtained manually according to the clinical criterion by using the ITK-SNAP tool \citep{yushkevich2006user}. And all subjects are provided after approval by the McGill University Health Centre. We perform one 5-fold cross-validation test to train our UAL for performance evaluation and comparison. Specifically, UAL is trained using batchsize of 2, iteration number of 100,000, and learning rate of 1e-4. The UAL is performed on {\itshape Ubuntu 18.04} platform, {\itshape Python v3.6}, {\itshape Pytorch v0.4.0}, and {\itshape CUDA v9.0} library, and running on {\itshape Intel(R) Core(TM) i9-9900K CPU @ 3.60GHz} and {\itshape GeForce GTX 1080Ti 11GB}.

\begin{figure*}[t]
\includegraphics[width=\textwidth]{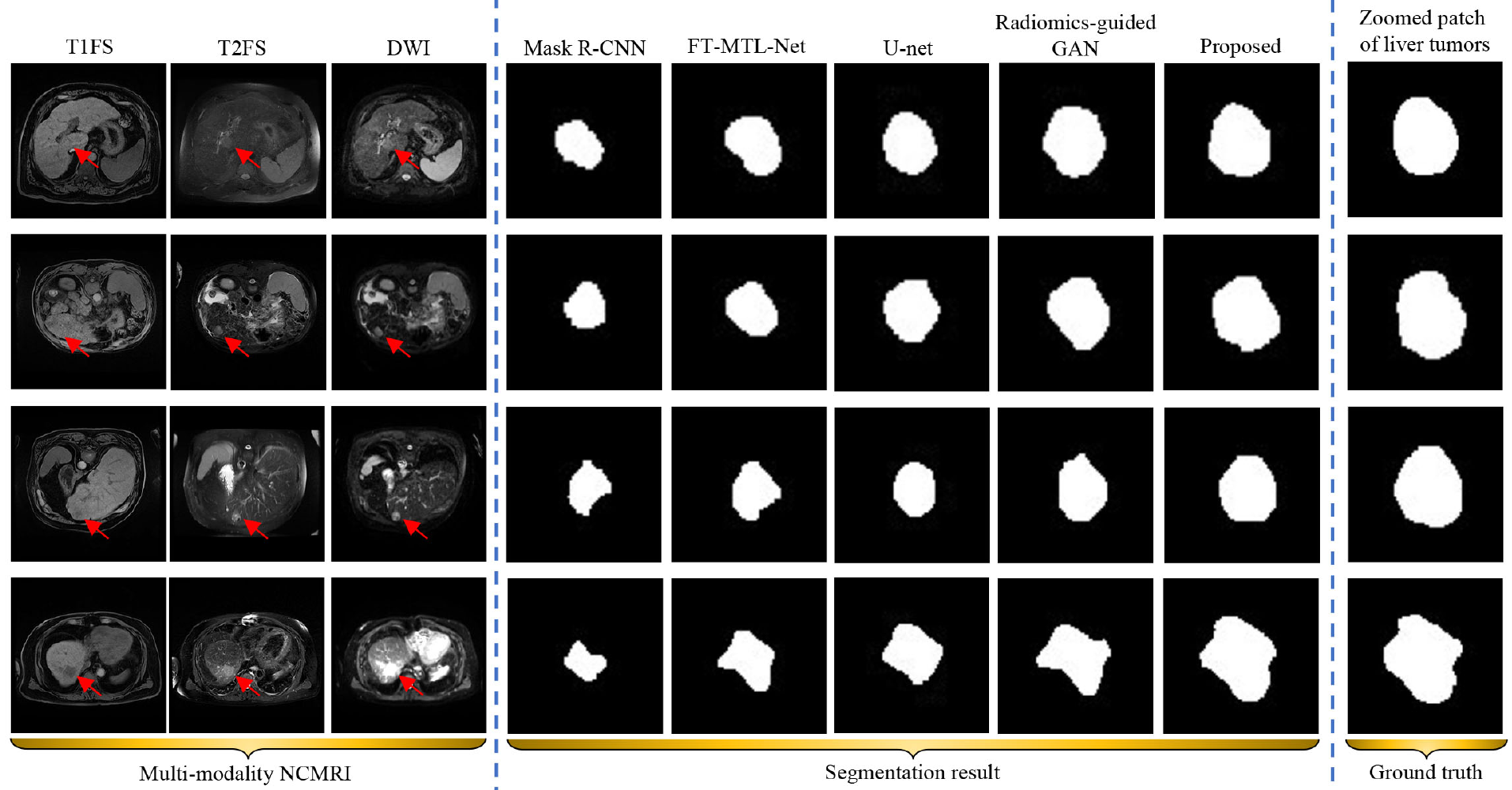}
\caption{The visual examples of the liver tumors segmentation, where the white denotes the liver tumors segmentations and the black denotes the background. From the left to right: source multi-modality NCMRI, segmentation results, and ground truth.} \label{Result}
\end{figure*}

\subsection{Evaluation metrics and method}
\subsubsection{Evaluation metrics} To quantitatively evaluate the segmentation performance of UAL, we utilize the DSC to calculate the similarity of the output ${\mathcal{Y}_n}$ and ground truth ${\hat{\mathcal{Y}}_n}$, which can be defined as:

\begin{equation}
DSC =\frac{1}{N}\sum_n^N\frac{2|{\mathcal{Y}_n} \cap {\hat{\mathcal{Y}}}_n|}{|{\mathcal{Y}_n}|+|{\hat{\mathcal{Y}}}_n|}\times100\%
\end{equation}
Using p-Acc to evaluate the proportion of pixels that are correctly classified. Using IoU to evaluate b-box regression performance via measuring accuracy of the output ${\mathcal{Y}_B}$ ($\mathcal{Y}_{Sc}$ before padding) relative to ground truth ${\hat{\mathcal{Y}}_B}$ (${\hat{\mathcal{Y}}}_{Sc}$ before padding). The IoU can be defined as:

\begin{equation}
DSC =\frac{1}{N}\sum_n^N\frac{2|{\mathcal{Y}_n} \cap {\hat{\mathcal{Y}}}_n|}{|{\mathcal{Y}_n}|+|{\hat{\mathcal{Y}}}_n|}\times100\%
\end{equation}
And we use the sensitivity (true positive rate, (TPR)), specificity (true negative rate, (TNR)), and accuracy (Acc) for evaluating liver tumors classification, in which the TPR, TNR, and Acc can be defined as:

\begin{equation}
TPR = \frac{TP}{TP+FN}\times100\%
\end{equation}
\begin{equation}
TNR = \frac{TN}{FP+TN}\times100\%
\end{equation}
\begin{equation}
Acc = \frac{TP+TN}{TP+FP+TN+FN}\times100\%
\end{equation}
where the hemangioma is defined as positive and HCC as negative. TP, FP, TN, and FN denotes the true positive, false positive, true negative, and false negative measurements, respectively.

\begin{table*}[t]
\centering
\caption{The quantitative evaluation of the liver tumors segmentation and detection. Six criteria (i.e. DSC, p-Acc, IoU, TPR, TNR, and Acc) evaluated the performance of our UAL and other six state-of-the-art methods. It demonstrates that our proposed UAL outperforms six state-of-the-art methods in liver tumors segmentation and detection.}\label{comparison}
\begin{tabular}{ ccccccc }
\hline
\itshape{Method} & DSC &p-Acc&IoU & TPR & TNR & Acc\\
\hline
\ \  \multirow{2}*{U-net}  &78.88 & 96.57& \multirow{2}*{-} &\multirow{2}*{-}  & \multirow{2}*{-} & \multirow{2}*{-}\\
\ \  &$\pm$3.73&$\pm$0.91&&&& \\
\cdashline{1-7}[0.8pt/2pt]
\ \ Radiomics-guided  &80.65&96.72& - & - & - & -\\
\ \  GAN&$\pm$3.13&$\pm$0.89&&&& \\
\cdashline{1-7}[0.8pt/2pt]
\ \ Faster  & \multirow{2}*{-} & \multirow{2}*{-} &  66.43 &78.63 &82.26 & 80.39\\
\ \  R-CNN&&&$\pm$8.20&$\pm$2.74&$\pm$2.45&$\pm$2.47 \\
\cdashline{1-7}[0.8pt/2pt]
\ \ \multirow{2}*{Tripartite-GAN}  & \multirow{2}*{-} & \multirow{2}*{-} & 73.42 &  86.82  &89.68 &88.24 \\
\ \  &&&$\pm$4.60&$\pm$1.84&$\pm$1.67& $\pm$1.78\\
\cdashline{1-7}[0.8pt/2pt]
\ \ Mask  &75.17&96.21&68.30&80.00 &83.20 &81.57 \\
\ \  R-CNN&$\pm$5.58&$\pm$1.16&$\pm$7.31&$\pm$2.43&$\pm$1.88&$\pm$2.09 \\
\cdashline{1-7}[0.8pt/2pt]
\ \ \multirow{2}*{FT-MTL-Net}  &77.58&96.48&70.64 &82.75&81.40 &84.13 \\
\ \  &$\pm$4.17&$\pm$0.93&$\pm$5.85&$\pm$2.20&$\pm$2.04&$\pm$2.11 \\
\hline
\ \ \multirow{2}*{\textbf{Proposed UAL}} &\textbf{83.63}&\textbf{97.75}&\textbf{81.30} &\textbf{92.13}&\textbf{93.75}&\textbf{92.94}\\
\ \ &\textbf{$\pm$2.16}&\textbf{$\pm$0.72}&\textbf{$\pm$3.26}&\textbf{$\pm$1.26}&\textbf{$\pm$0.74}&\textbf{$\pm$0.86} \\
\hline
\end{tabular}
\end{table*}

\subsection{Performance comparison with state-of-the-art}
The UAL has been validated by comparing with two state-of-the-art segmentation methods (U-net \citep{ronneberger2015u} and Radiomics-guided GAN \citep{xiao2019radiomics}), two state-of-the-art detection methods (Faster R-CNN \citep{ren2015faster} and Tripartite-GAN \citep{zhao2020tripartite}), and two state-of-the-art simultaneous segmentation and detection methods (Mask R-CNN \citep{he2017mask} and FT-MTL-Net\citep{gao2020feature}). The visual segmentation results are shown in Fig.\ref{Result}. The quantitative analysis results of segmentation and detection are shown in Table.\ref{comparison}. Our UAL outperforms six state-of-the-art methods, which achieved liver tumors segmentation with DSC of 83.63\% and p-Acc of 97.75\%, b-box regression with IoU of 81.30\%, and liver tumors classification with TPR of 92.13\%, TNR of 93.75\%, and Acc of 92.94\%. These high performances in both the segmentation and the detection are from EDFPM, FSC, CSWP, and MPRG-D.

\begin{figure*}[ht]
\includegraphics[width=\textwidth]{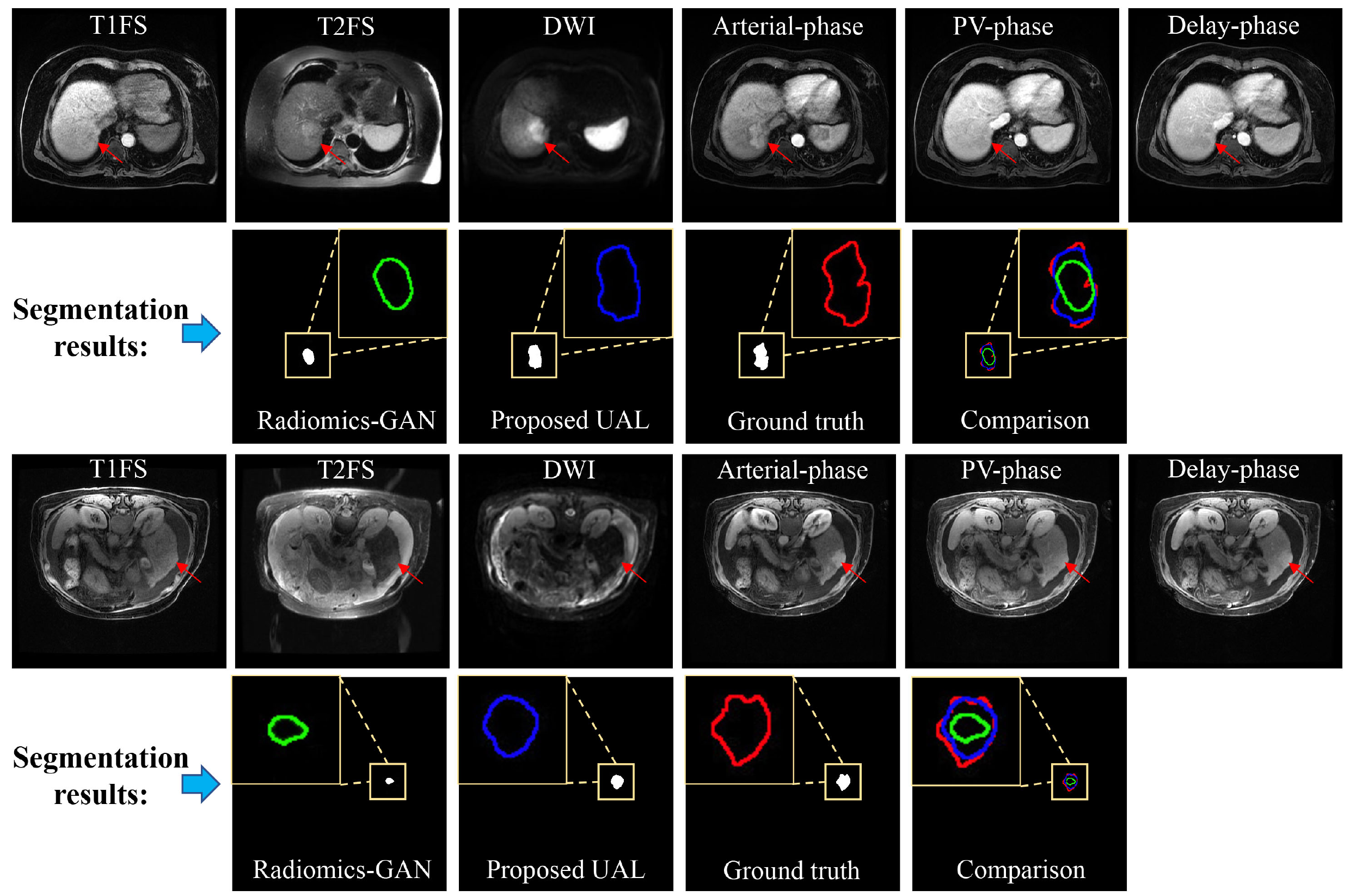}
\caption{Two cases of liver tumors information on T2FS is invisible or insufficient, and the liver tumors information on arterial-phase CEMRI is more clear than delay-phase CEMRI. The first and third rows are multi-modality NCMRI and multi-phase CEMRI. From the left to right: T1FS, T2FS, DWI, arterial-phase CEMRI, PV-phase CEMRI, and delay-phase CEMRI. The second and fourth rows are segmentation results. From the left to right: result from Radiomics-GAN, result from our UAL, ground truth, and the comparison of segmentation results. It is clear that our proposed UAL outperforms the Radiomics-GAN.} 
\label{Result2}
\end{figure*}

Moreover, to validate the contributions of using multi-modality NCMRI and multi-modality CEMRI, we showed the cases of the liver tumors information are invisible or insufficient on T2FS. And the liver tumors information is more clear on arterial-phase than delay-phase CEMRI. The comparisons are performed among Radiomics-GAN \citep{xiao2019radiomics}  and our UAL, in which the Radiomics-GAN uses T2FS and delay-phase CEMRI for liver tumors segmentation. The visualized results are shown in Fig.\ref{Result2}. It is clear that our proposed UAL outperforms Radiomics-GAN. The results demonstrate that our UAL has high robustness via using multi-modality NCMRI and multi-modality CEMRI than Radiomics-GAN. Especially for the situation that liver tumors information is invisible or insufficient on T2FS, and the liver tumors information is more clear on arterial-phase than delay-phase CEMRI.

\begin{table*}[t]
\centering
\caption{The quantitative evaluation of the ablation studies. The ablation studies demonstrate that every part of the newly designed UAL contributes to liver tumors segmentation and detection.}\label{ablation}
\begin{tabular}{ ccccccc }
\hline
\itshape{Method} & DSC &p-Acc&IoU & TPR & TNR & Acc\\
\hline
\ \ Without&81.65&96.88&76.77&91.34 &92.97 &92.61 \\
\ \ EDFPM&$\pm$2.76&$\pm$0.81&$\pm$3.85&$\pm$1.56&$\pm$1.31&$\pm$1.28 \\
\cdashline{1-7}[0.8pt/2pt]
\ \ Without &81.37&96.86&76.29&90.55 &92.19 &91.37 \\
\ \ FSC&$\pm$2.92&$\pm$0.85&$\pm$4.13&$\pm$1.60&$\pm$1.42&$\pm$1.36 \\
\cdashline{1-7}[0.8pt/2pt]
\ \ Without &82.33&97.12&74.82&89.76 &91.41 &90.59 \\
\ \ CSWP&$\pm$2.46&$\pm$0.78&$\pm$4.53&$\pm$1.66&$\pm$1.51&$\pm$1.45 \\
\cdashline{1-7}[0.8pt/2pt]
\ \ Without &80.67&96.27&74.47&84.50 &87.30 &85.88\\
\ \ MPR&$\pm$3.06&$\pm$0.88&$\pm$4.32&$\pm$1.93&$\pm$1.76&$\pm$1.82  \\
\cdashline{1-7}[0.8pt/2pt]
\ \ Without &80.53&96.19&73.37&83.72 &86.51 &85.10 \\
\ \ MPRG-D&$\pm$2.96&$\pm$0.89&$\pm$4.56&$\pm$2.01&$\pm$1.77&$\pm$1.88 \\
\cdashline{1-7}[0.8pt/2pt]
\ \ \multirow{2}*{\textbf{Proposed UAL}} &\textbf{83.63}&\textbf{97.75}&\textbf{81.30} &\textbf{92.13}&\textbf{93.75}&\textbf{92.94}\\
\ \ &\textbf{$\pm$2.16}&\textbf{$\pm$0.72}&\textbf{$\pm$3.26}&\textbf{$\pm$1.26}&\textbf{$\pm$0.74}&\textbf{$\pm$0.86} \\
\hline
\end{tabular}
\end{table*}

\subsection{Ablation studies}
In order to verify the contributions of EDFPM, FSC, CSWP, MPR, and MPRG-D. We performed the comparison among our UAL, the UAL without EDFPM, the UAL without FSC, the UAL without CSWP, the UAL without MPR, and the UAL without MPRG-D. The quantitative analysis results of these ablation studies are shown in Table.\ref{ablation}, which demonstrated that every part of the UAL contributes to the liver tumors segmentation and detection.

\subsubsection{Evaluation of EDFPM}
\begin{figure}[h]
\includegraphics[width=1\textwidth]{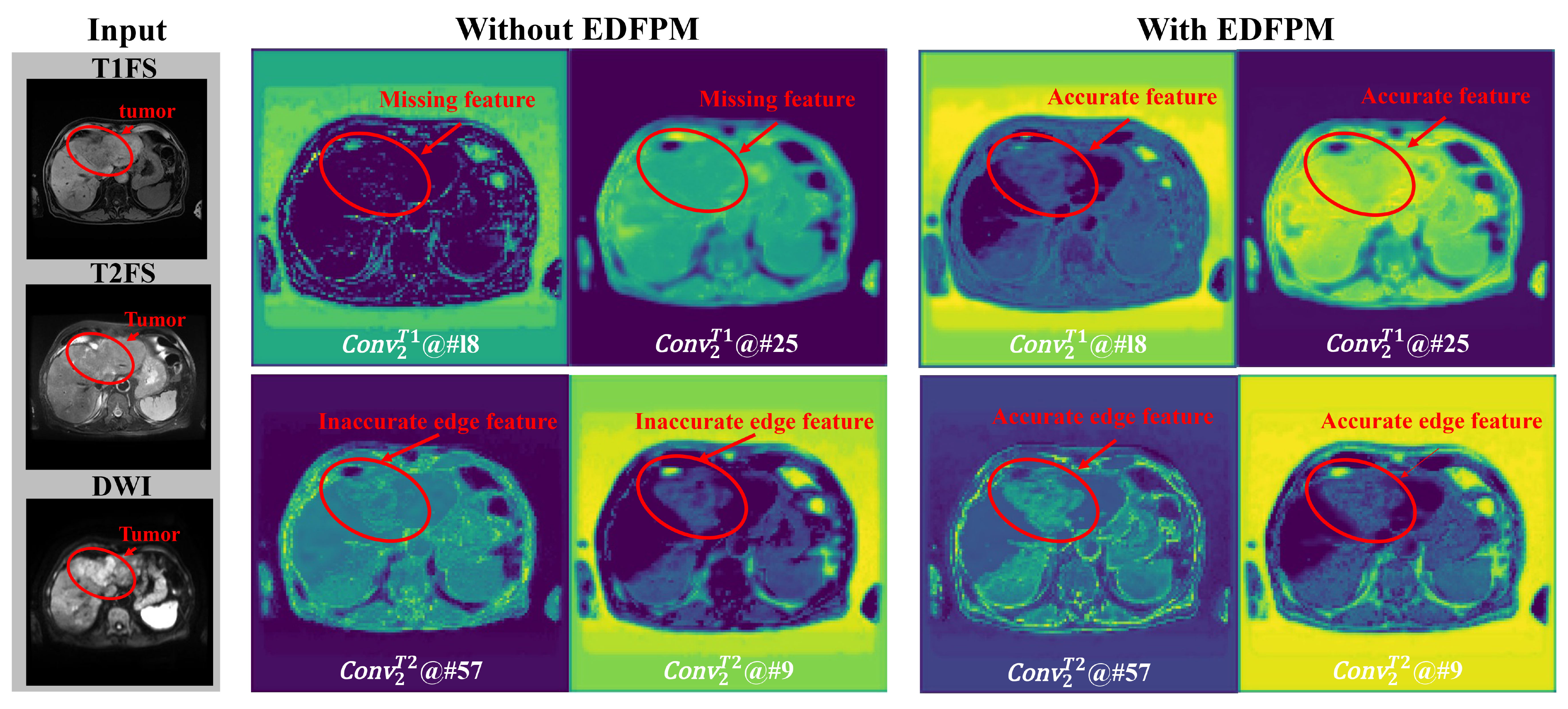}
\caption{The comparison of feature maps from proposed UAL and UAL without EDFPM. It is clear that EDFPM contributes to accurate edge features and HCC features extraction, which are marked out in the red circle.} \label{EDFPM}
\end{figure}
To verify the contribution of EDFPM, we performed the comparison between the UAL and the UAL without EDFPM. The quantitative results ($2^{nd}$ row in Table.\ref{ablation}) showed that the performance of segmentation and detection dropped when EDFPM is removed. Specifically, the DSC decreased by 1.98\%, the IoU decreased by 4.53\%, and the Acc decreased by 0.33\%. It demonstrates that the newly designed EDFPM benefits liver tumors segmentation and detection. Besides, in order to visualize the contribution of that EDFPM facilitates the complementary multi-modality NCMRI information extraction, we showed the visualized feature maps in Fig.\ref{EDFPM}. For the first row of feature maps, they are obtained from the channel \#18 and channel \#25 of $Conv_2^{T1}$ layer. It is clear that the feature maps lost the feature of the liver tumors when without EDFPM. In contrast, the feature maps obtained by our proposed UAL are precise. This demonstrates that the complementary feature between T1FS and T2FS extracted by EDFPM facilitates the liver tumors feature extraction. For the second row of feature maps, they are obtained from the channel \#57 and channel \#9 of $Conv_2^{T2}$ layer. It is clear that the edge feature of liver tumors in feature maps obtained without EDFPM is inaccurate. In contrast, the feature maps obtained by our proposed UAL are precise. This demonstrates that the complementary feature between T2FS and DWI extracted by EDFPM facilitates the liver tumors edge feature extraction. To summarize, the EDFPM extracts the multi-size edge dissimilarity maps, which as the prior knowledge added into the three parallel convolution channels benefits to multi-modality NCMRI feature extraction for improving liver tumors segmentation and detection.

\subsubsection{Evaluation of FSC}
To verify the contribution of FSC, we use the operation of concatenation followed by the convolution and ReLU to replace the FSC (i.e. without FSC). It means that using the $\mathcal{F}^{s}=\varepsilon(x^{[\mathcal{F}^{T1}, \mathcal{F}^{T2}, \mathcal{F}^{D}]}*\mathcal{W}_s+b_s)$ to replace $\mathcal{F}_{Seg}$ and $\mathcal{F}_{Dec}$. The quantitative results ($3^{rd}$ row in Table.\ref{ablation}) showed that the performance of segmentation and detection dropped when FSC is replaced. In which the DSC decreased by 2.26\%, the IoU decreased by 5.01\%, and the Acc decreased by 1.57\%. It demonstrates that the design with the gate and the weight increasing of $\mathcal{F}^{T1}$ and $\mathcal{F}^{D}$ contributes the liver tumors segmentation and detection.

\begin{figure}[h]
\includegraphics[width=0.9\textwidth]{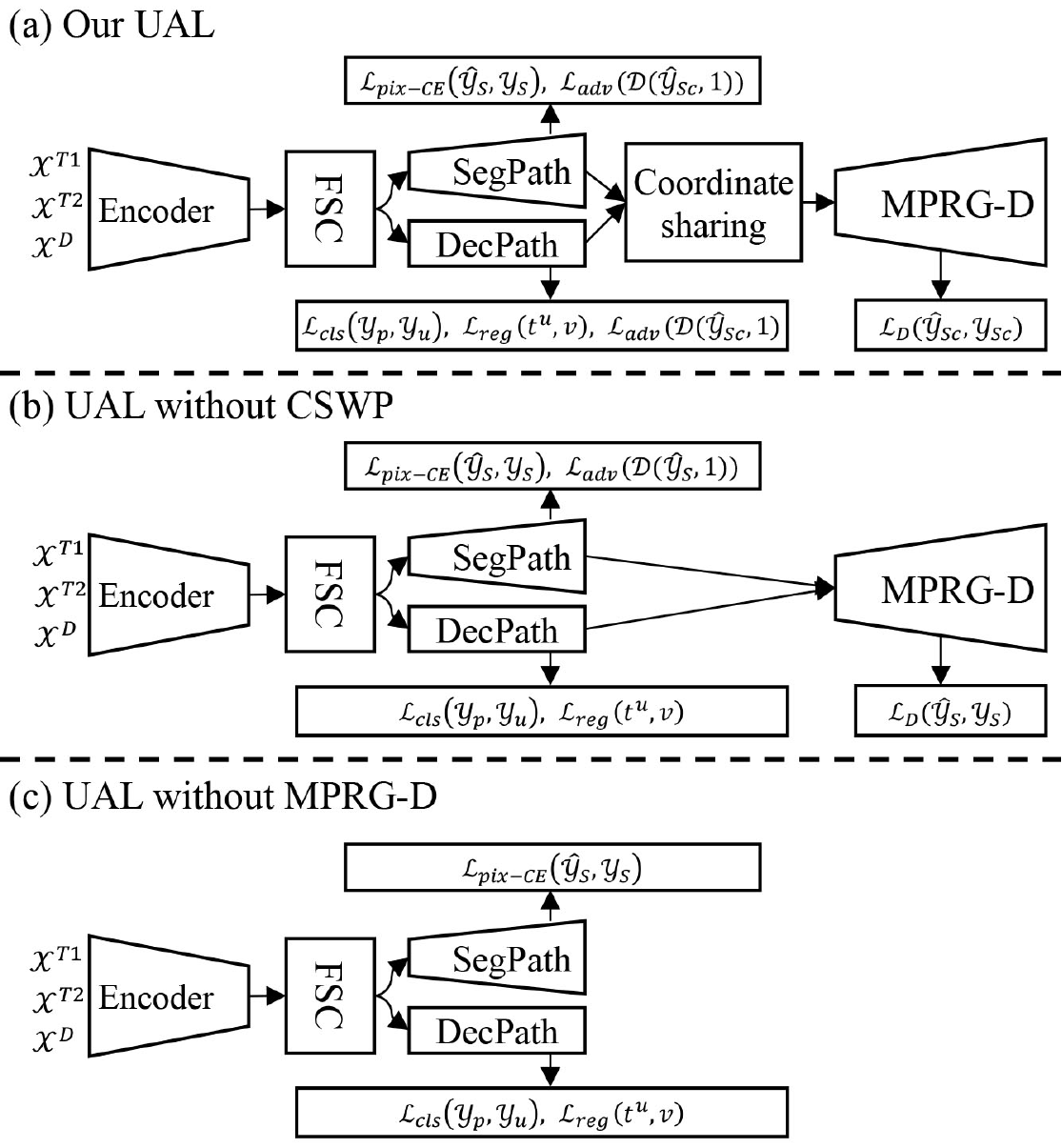}
\caption{Structures of the (a) our proposed UAL, (b) UAL without CSWP, and (c) UAL without MPRG-D.} \label{Structure}
\end{figure}

\subsubsection{Evaluation of CSWP}
To verify the contribution of CSWP, we performed the comparison between the UAL and the UAL without CSWP. The structures of the UAL and the UAL without CSWP are shown in Fig.\ref{Structure} (a) and (b). Fig.\ref{Structure}(a) indicates that CSWP enables the SegPath (i.e. segmentation path) and DecPath (i.e. detection path) to perform the united adversarial learning by $\mathcal{L}_{adv}(\mathcal{D}({\hat{\mathcal{Y}}}_{Sc}, 1))$. Fig.\ref{Structure}(b) indicates that when CSWP removed, it only enables the SegPath to perform adversarial learning by $\mathcal{L}_{adv}(\mathcal{D}({\hat{\mathcal{Y}}}_{S}, 1))$ and no adversarial learning in DecPath. The quantitative result ($6^{th}$ row in Table.\ref{ablation}) showed that the performance of segmentation and detection dropped when CSWP is removed. Especially for the b-box regression, the IoU value decreased from 81.30\% to 74.82\% (decreased by 6.48\%). This proved that CSWP enables segmentation and detection can be unified to perform adversarial learning for improving their performance. Especially for the great performance improvement of b-box regression by using the united adversarial learning strategy.

\begin{figure}[h]
\includegraphics[width=1\textwidth]{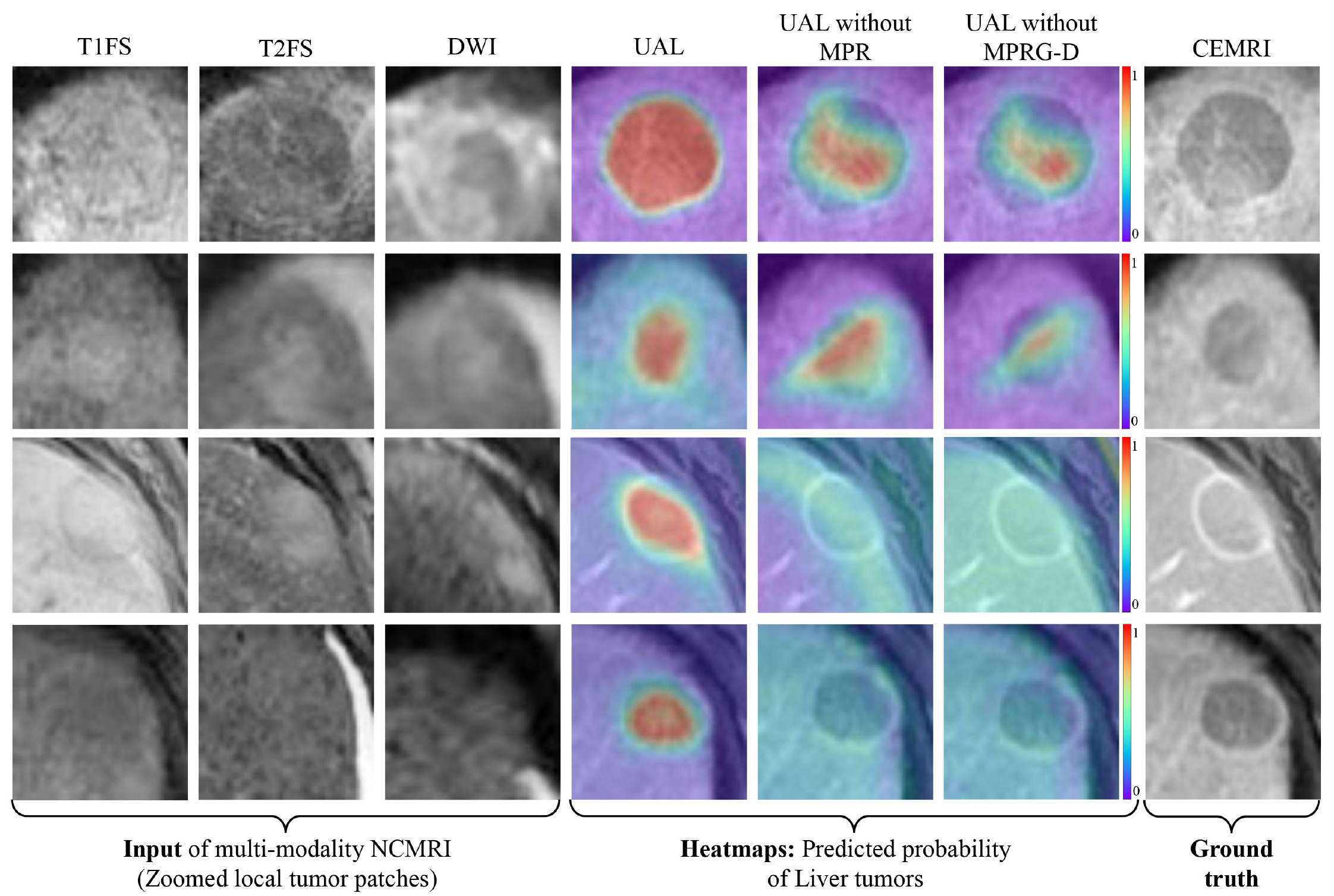}
\caption{Four cases of the comparison of heatmaps (predicted probability of liver tumors, highlighted region by red color means high predicted score after $softmax$ layer) from proposed UAL, UAL without MPR, and UAL without MPRG-D. From the left to right, it is T1FS, T2FS, DWI, heatmaps from UAL, heatmaps from UAL without MPR, heatmaps from UAL without MPRG-D, and ground truth of liver tumors from CEMRI, respectively.} \label{heatmap}
\end{figure}

\subsubsection{Evaluation of MPR and MPRG-D}
To verify the contribution of MPRG-D, we removed the MPRG-D in UAL, which is shown in Fig.\ref{Structure}(c). In this case, neither SegPath nor DecPath has adversarial learning. The performance of segmentation and detection dropped significantly ($6^{th}$ row in Table.\ref{ablation}), in which the DSC decreased by 3.10\%, the IoU decreased by 7.93\%, and the Acc decreased by 7.84\%. To further verify the contribution of MPR in MPRG-D, we keep adversarial learning but remove the MPR. The quantitative results ($5^{th}$ row in Table.\ref{ablation}) showed that the performance of segmentation and detection still dropped significantly, in which the DSC decreased by 2.96\%, the IoU decreased by 6.83\%, and the Acc decreased by 7.06\%. This shows that adversarial learning is greatly helpful for liver tumors segmentation and detection improvement. And most of the contribution comes from MPR. In order to visualize the contribution, we show the four cases heatmaps from UAL, UAL without MPR, and UAL without MPRG-D in Fig.\ref{heatmap}. From the first two rows in Fig.\ref{heatmap}, it is clear that the predicted scores of tumor area obtained from our UAL are more accurate than UAL without MPR and UAL without MPRG-D. From the last two rows in Fig.\ref{heatmap}, it is clear that the UAL without MPR and UAL without MPRG-D lose tumors prediction when liver tumors are invisible on multi-modality NCMRI.  All these results demonstrate that MPR and MPRG-D improved UAL performance when liver tumors information is invisible or insufficient on multi-modality NCMRI.

\subsection{Influences of the combination of different NCMRI modality}
In order to verify the influences of the combination of different NCMRI modality, we set up six different combinations of different NCMRI modality as the inputs of our UAL. Quantitative analysis results are shown in Table.\ref{NCMRI}. Firstly, when the input of UAL is a single modality of NCMRI (i.e. T1FS only, T2FS only, or DWI only), the quantitative results ($3^{rd}$, $4^{th}$, and $5^{th}$ rows in Table.\ref{NCMRI}) show that the performance of segmentation and detection among these three single modalities are roughly equivalent, except for the IoU value ($6^{th}$ column in Table.\ref{NCMRI}). Compare with DWI only, the IoU value decreased by 2.93\% of using T1FS only and decreased by 2.48\% of using T2FS only. This demonstrated that DWI is helpful to detect the location of liver tumors when using a single modality of NCMRI. Secondly, when the input of UAL is the combination of two NCMRI modalities, the quantitative results ($6^{th}$, $7^{th}$, and $8^{th}$ rows in Table.\ref{NCMRI}) show the performance of segmentation and detection among these combinations are roughly equivalent. And they are all better than using sing modality only, which demonstrates that the complementary information in different NCMRI modalities benefits liver tumors segmentation and detection. Lastly, the quantitative results of the last row in Table.\ref{NCMRI} of using multi-modality NCMRI (i.e. T1FS, T2FS, and DWI) achieves the best results compared with the other six different combinations of multi-modality NCMRI. It proved that using multi-modality NCMRI information is the best choice for our UAL to perform liver tumors segmentation and detection.

\begin{table*}[t]
\centering
\caption{The quantitative evaluation of 6 types of combinations of multi-modality NCMRI for liver tumors segmentation and detection. It demonstrates that using multi-modality NCMRI information is the best choice for our UAL to perform liver tumors segmentation and detection.}\label{NCMRI}
\begin{tabular}{ ccccccccc }
\hline
 \multicolumn{3}{c}{Modality of NCMRI}& \multirow{2}*{DSC} &\multirow{2}*{p-Acc}&\multirow{2}*{IoU}&\multirow{2}*{TPR}&\multirow{2}*{TNR}&\multirow{2}*{Acc} \\
 \cline{1-3}
T1FS& T2FS& DWI&&&&&&\\
\hline
\multirow{2}*{\checkmark} &&&81.09&96.05&75.33&  87.40 & 90.03  &88.71   \\
&&&$\pm$2.58&$\pm$0.87&$\pm$3.98&$\pm$1.77&$\pm$1.65&$\pm$1.65\\
\cdashline{1-7}[0.8pt/2pt]
&\multirow{2}*{\checkmark} &&81.16&96.13&75.78&  87.11 &  89.86 &88.47  \\
&&&$\pm$2.57&$\pm$0.86&$\pm$3.77&$\pm$1.80&$\pm$1.68&$\pm$1.70\\
\cdashline{1-7}[0.8pt/2pt]
&&\multirow{2}*{\checkmark} & 81.54&96.51&78.26& 87.56 &  90.19 & 88.86 \\
&&&$\pm$2.65&$\pm$0.83&$\pm$3.37&$\pm$1.72&$\pm$1.60&$\pm$1.68\\
\cdashline{1-7}[0.8pt/2pt]
\multirow{2}*{\checkmark}&\multirow{2}*{\checkmark}&&81.58&96.76&76.42& 87.71 & 90.35 &89.02 \\
&&&$\pm$2.54&$\pm$0.82&$\pm$3.69&$\pm$1.69&$\pm$1.58&$\pm$1.58\\
\cdashline{1-7}[0.8pt/2pt]
\multirow{2}*{\checkmark}& &\multirow{2}*{\checkmark}&82.89&97.16& 79.96  & 88.58& 90.72 & 89.65 \\
&&&$\pm$2.30&$\pm$0.78&$\pm$3.32&$\pm$1.67&$\pm$1.57&$\pm$1.55\\
\cdashline{1-7}[0.8pt/2pt]
&\multirow{2}*{\checkmark}&\multirow{2}*{\checkmark}&82.67&96.92&78.14&   89.31 &  91.08 &  90.20\\
&&&$\pm$2.35&$\pm$0.77&$\pm$3.41 &$\pm$1.58&$\pm$1.53&$\pm$1.49\\
\cdashline{1-7}[0.8pt/2pt]
\textbf{\multirow{2}*{\checkmark} }& \textbf{\multirow{2}*{\checkmark}}&\textbf{\multirow{2}*{\checkmark}}&\textbf{83.63}&\textbf{97.75}&\textbf{81.30} &\textbf{92.13}&\textbf{93.75}&\textbf{92.94}\\
&&&\textbf{$\pm$2.16}&\textbf{$\pm$0.72}&\textbf{$\pm$3.26}&\textbf{$\pm$1.26}&\textbf{$\pm$0.74}&\textbf{$\pm$0.86}\\
\hline
\end{tabular}
\end{table*}

\subsection{Influences of the combination of different CEMRI phase}
In order to verify the influences of the combinations of different CEMRI phase, we set up six different combinations of different CEMRI phase for radiomics feature extraction in our MPRG-D. Quantitative analysis results are shown in Table.\ref{CEMRI}. Firstly, when we using a single phase CEMRI in our MPRG-D (i.e. arterial-phase only, PV-phase only, or delay-phase only), the quantitative results ($3^{rd}$, $4^{th}$, and $5^{th}$ rows in Table.\ref{CEMRI}) illustrates that the delay-phase achieves the best performance. This proved that the radiomics feature extracted from delay-phase CEMRI is the most useful for liver tumors segmentation and detection than Arterial-phase CEMRI and PV-phase CEMRI. Secondly, when we using three combinations of two phases CEMRI, the quantitative results ($6^{th}$, $7^{th}$, and $8^{th}$ rows in Table.\ref{CEMRI}) show that the performance of segmentation and detection is better than using a single phase CEMRI. Lastly, the quantitative results of the last row in Table.\ref{CEMRI} of using multi-phases CEMRI (i.e. arterial-phase, PV-phase, or delay-phase) achieves the best results compared with the other six different combinations of multi-phase CEMRI. All these results proved that extracting multi-phase CEMRI radiomics feature maximizes the ability of MPRG-D to discriminate. And then improve the performance of liver tumors segmentation and detection via using the united adversarial learning strategy.

\begin{table*}[t]
\centering
\caption{The quantitative evaluation of 6 types of combinations of multi-phase CEMRI for liver tumors segmentation and detection. It demonstrates that using multi-phase CEMRI information to perform the united adversarial learning strategy is the best choice for our UAL.}\label{CEMRI}
\begin{tabular}{ ccccccccc }
\hline
 \multicolumn{3}{c}{Phase of CEMRI}& \multirow{2}*{DSC} &\multirow{2}*{p-Acc}&\multirow{2}*{IoU}&\multirow{2}*{TPR}&\multirow{2}*{TNR}&\multirow{2}*{Acc} \\
 \cline{1-3}
Arterial& PV & Delay &&&&&&\\
\hline
\multirow{2}*{\checkmark} &&&81.43&96.36&79.77&  91.02& 92.65  & 91.83   \\
&&&$\pm$2.55&$\pm$0.85&$\pm$3.50&$\pm$1.56 &$\pm$1.15&$\pm$1.28\\
\cdashline{1-8}[0.8pt/2pt]
&\multirow{2}*{\checkmark} &&80.92&96.03&79.33&  90.71 &  92.34 &91.53  \\
&&&$\pm$2.63&$\pm$0.86&$\pm$3.63&$\pm$1.62&$\pm$1.31&$\pm$1.38\\
\cdashline{1-8}[0.8pt/2pt]
&&\multirow{2}*{\checkmark} & 81.80&96.86&79.85& 91.18 &  92.81 & 92.00 \\
&&&$\pm$2.45&$\pm$0.83&$\pm$3.54&$\pm$1.55&$\pm$1.04&$\pm$1.23\\
\cdashline{1-8}[0.8pt/2pt]
\multirow{2}*{\checkmark}&\multirow{2}*{\checkmark}&&82.18&96.90&79.91&91.50 & 93.13 & 92.31 \\
&&&$\pm$2.44&$\pm$0.81&$\pm$3.55&$\pm$1.44&$\pm$0.85&$\pm$1.10\\
\cdashline{1-8}[0.8pt/2pt]
\multirow{2}*{\checkmark}& &\multirow{2}*{\checkmark}&83.19&97.16& 81.13  & 91.81&  93.44 & 92.63 \\
&&&$\pm$2.20&$\pm$0.76&$\pm$3.30&$\pm$1.39&$\pm$0.82&$\pm$1.07\\
\cdashline{1-8}[0.8pt/2pt]
&\multirow{2}*{\checkmark}&\multirow{2}*{\checkmark}&83.06&96.98&81.05 &   91.65& 93.28 & 92.47 \\
&&&$\pm$2.27&$\pm$0.76&$\pm$3.36&$\pm$1.43&$\pm$0.88&$\pm$1.15\\
\cdashline{1-8}[0.8pt/2pt]
\textbf{\multirow{2}*{\checkmark} }& \textbf{\multirow{2}*{\checkmark}}&\textbf{\multirow{2}*{\checkmark}}&\textbf{83.63}&\textbf{97.75}&\textbf{81.30} &\textbf{92.13}&\textbf{93.75}&\textbf{92.94}\\
&&&\textbf{$\pm$2.16}&\textbf{$\pm$0.72}&\textbf{$\pm$3.26}&\textbf{$\pm$1.26}&\textbf{$\pm$0.74}&\textbf{$\pm$0.86}\\
\hline
\end{tabular}
\end{table*}

\section{Conclusions}
For the first time, the proposed UAL achieves simultaneous segmentation and detection of HCC via using multi-modality NCMRI only. The novel EDFPM extracts the multi-size edge dissimilarity maps that enhance multi-modality NCMRI information extraction. And then the innovative FSC fuses the multi-modality NCMRI feature and makes the final decision of feature selection according to the liver tumors segmentation and detection. Finally, the newly designed MPRG-D enhances discrimination by adding the MPR feature. And with the help of the proposed CSWP mechanism, the MPRG-D achieves united adversarial learning for promoting liver tumors segmentation and detection. The experimental results (i.e. DSC of 83.63\%, p-Acc of 97.75\%, IoU of 81.30\%, the sensitivity of 92.13\%, the specificity of 93.75\%, and detection accuracy of 92.94\%) demonstrate that UAL has great potential to assist clinical segmentation and detection of liver tumors without CAs injection. 

\bibliographystyle{model2-names}\biboptions{authoryear}
\bibliography{mybibfile}

\begin{thebibliography}{34}
\expandafter\ifx\csname natexlab\endcsname\relax\def\natexlab#1{#1}\fi
\providecommand{\url}[1]{\texttt{#1}}
\providecommand{\href}[2]{#2}
\providecommand{\path}[1]{#1}
\providecommand{\DOIprefix}{doi:}
\providecommand{\ArXivprefix}{arXiv:}
\providecommand{\URLprefix}{URL: }
\providecommand{\Pubmedprefix}{pmid:}
\providecommand{\doi}[1]{\href{http://dx.doi.org/#1}{\path{#1}}}
\providecommand{\Pubmed}[1]{\href{pmid:#1}{\path{#1}}}
\providecommand{\bibinfo}[2]{#2}
\ifx\xfnm\relax \def\xfnm[#1]{\unskip,\space#1}\fi
\bibitem[{Bousabarah et~al.(2020)Bousabarah, Letzen, Tefera, Savic, Schobert,
  Schlachter, Staib, Kocher, Chapiro and Lin}]{bousabarah2020automated}
\bibinfo{author}{Bousabarah, K.}, \bibinfo{author}{Letzen, B.},
  \bibinfo{author}{Tefera, J.}, \bibinfo{author}{Savic, L.},
  \bibinfo{author}{Schobert, I.}, \bibinfo{author}{Schlachter, T.},
  \bibinfo{author}{Staib, L.H.}, \bibinfo{author}{Kocher, M.},
  \bibinfo{author}{Chapiro, J.}, \bibinfo{author}{Lin, M.},
  \bibinfo{year}{2020}.
\newblock \bibinfo{title}{Automated detection and delineation of hepatocellular
  carcinoma on multiphasic contrast-enhanced mri using deep learning}.
\newblock \bibinfo{journal}{Abdominal Radiology (New York)} .
\bibitem[{Canellas et~al.(2019)Canellas, Patel, Agarwal and
  Sahani}]{canellas2019lesion}
\bibinfo{author}{Canellas, R.}, \bibinfo{author}{Patel, M.J.},
  \bibinfo{author}{Agarwal, S.}, \bibinfo{author}{Sahani, D.V.},
  \bibinfo{year}{2019}.
\newblock \bibinfo{title}{Lesion detection performance of an abbreviated
  gadoxetic acid--enhanced mri protocol for colorectal liver metastasis
  surveillance}.
\newblock \bibinfo{journal}{European radiology} \bibinfo{volume}{29},
  \bibinfo{pages}{5852--5860}.
\bibitem[{Cereser et~al.(2010)Cereser, Furlan, Bagatto, Girometti, Como,
  Avellini, Orsaria, Zuiani and Bazzocchi}]{cereser2010comparison}
\bibinfo{author}{Cereser, L.}, \bibinfo{author}{Furlan, A.},
  \bibinfo{author}{Bagatto, D.}, \bibinfo{author}{Girometti, R.},
  \bibinfo{author}{Como, G.}, \bibinfo{author}{Avellini, C.},
  \bibinfo{author}{Orsaria, M.}, \bibinfo{author}{Zuiani, C.},
  \bibinfo{author}{Bazzocchi, M.}, \bibinfo{year}{2010}.
\newblock \bibinfo{title}{Comparison of portal venous and delayed phases of
  gadolinium-enhanced magnetic resonance imaging study of cirrhotic liver for
  the detection of contrast washout of hypervascular hepatocellular carcinoma}.
\newblock \bibinfo{journal}{Journal of computer assisted tomography}
  \bibinfo{volume}{34}, \bibinfo{pages}{706--711}.
\bibitem[{Choi et~al.(2014)Choi, Lee and Sirlin}]{choi2014ct}
\bibinfo{author}{Choi, J.Y.}, \bibinfo{author}{Lee, J.M.},
  \bibinfo{author}{Sirlin, C.B.}, \bibinfo{year}{2014}.
\newblock \bibinfo{title}{Ct and mr imaging diagnosis and staging of
  hepatocellular carcinoma: part ii. extracellular agents, hepatobiliary
  agents, and ancillary imaging features}.
\newblock \bibinfo{journal}{Radiology} \bibinfo{volume}{273},
  \bibinfo{pages}{30--50}.
\bibitem[{Ebeed et~al.(2017)Ebeed, Romeih, Refat and Yossef}]{ebeed2017role}
\bibinfo{author}{Ebeed, A.E.}, \bibinfo{author}{Romeih, M.A.E.h.},
  \bibinfo{author}{Refat, M.M.}, \bibinfo{author}{Yossef, M.H.},
  \bibinfo{year}{2017}.
\newblock \bibinfo{title}{Role of dynamic contrast-enhanced and diffusion
  weighted mri in evaluation of hepatocellular carcinoma after
  chemoembolization}.
\newblock \bibinfo{journal}{The Egyptian Journal of Radiology and Nuclear
  Medicine} \bibinfo{volume}{48}, \bibinfo{pages}{807--815}.
\bibitem[{Gao et~al.(2020)Gao, Yoon, Wu and Chu}]{gao2020feature}
\bibinfo{author}{Gao, F.}, \bibinfo{author}{Yoon, H.}, \bibinfo{author}{Wu,
  T.}, \bibinfo{author}{Chu, X.}, \bibinfo{year}{2020}.
\newblock \bibinfo{title}{A feature transfer enabled multi-task deep learning
  model on medical imaging}.
\newblock \bibinfo{journal}{Expert Systems with Applications}
  \bibinfo{volume}{143}, \bibinfo{pages}{112957}.
\bibitem[{Ge et~al.(2019)Ge, Yang, Chen, Luo, Feng, Ma, Ren and Li}]{ge2019k}
\bibinfo{author}{Ge, R.}, \bibinfo{author}{Yang, G.}, \bibinfo{author}{Chen,
  Y.}, \bibinfo{author}{Luo, L.}, \bibinfo{author}{Feng, C.},
  \bibinfo{author}{Ma, H.}, \bibinfo{author}{Ren, J.}, \bibinfo{author}{Li,
  S.}, \bibinfo{year}{2019}.
\newblock \bibinfo{title}{K-net: Integrate left ventricle segmentation and
  direct quantification of paired echo sequence}.
\newblock \bibinfo{journal}{IEEE transactions on medical imaging}
  \bibinfo{volume}{39}, \bibinfo{pages}{1690--1702}.
\bibitem[{Goodfellow et~al.(2014)Goodfellow, Pouget-Abadie, Mirza, Xu,
  Warde-Farley, Ozair, Courville and Bengio}]{goodfellow2014generative}
\bibinfo{author}{Goodfellow, I.}, \bibinfo{author}{Pouget-Abadie, J.},
  \bibinfo{author}{Mirza, M.}, \bibinfo{author}{Xu, B.},
  \bibinfo{author}{Warde-Farley, D.}, \bibinfo{author}{Ozair, S.},
  \bibinfo{author}{Courville, A.}, \bibinfo{author}{Bengio, Y.},
  \bibinfo{year}{2014}.
\newblock \bibinfo{title}{Generative adversarial nets}, in:
  \bibinfo{booktitle}{Advances in neural information processing systems}, pp.
  \bibinfo{pages}{2672--2680}.
\bibitem[{Hamm et~al.(2019)Hamm, Wang, Savic, Ferrante, Schobert, Schlachter,
  Lin, Duncan, Weinreb, Chapiro et~al.}]{hamm2019deep}
\bibinfo{author}{Hamm, C.A.}, \bibinfo{author}{Wang, C.J.},
  \bibinfo{author}{Savic, L.J.}, \bibinfo{author}{Ferrante, M.},
  \bibinfo{author}{Schobert, I.}, \bibinfo{author}{Schlachter, T.},
  \bibinfo{author}{Lin, M.}, \bibinfo{author}{Duncan, J.S.},
  \bibinfo{author}{Weinreb, J.C.}, \bibinfo{author}{Chapiro, J.}, et~al.,
  \bibinfo{year}{2019}.
\newblock \bibinfo{title}{Deep learning for liver tumor diagnosis part i:
  development of a convolutional neural network classifier for multi-phasic
  mri}.
\newblock \bibinfo{journal}{European radiology} \bibinfo{volume}{29},
  \bibinfo{pages}{3338--3347}.
\bibitem[{Han et~al.(2018)Han, Choi, Park, Choi, Rha and
  Lee}]{han2018diagnostic}
\bibinfo{author}{Han, S.}, \bibinfo{author}{Choi, J.I.}, \bibinfo{author}{Park,
  M.Y.}, \bibinfo{author}{Choi, M.H.}, \bibinfo{author}{Rha, S.E.},
  \bibinfo{author}{Lee, Y.J.}, \bibinfo{year}{2018}.
\newblock \bibinfo{title}{The diagnostic performance of liver mri without
  intravenous contrast for detecting hepatocellular carcinoma: a
  case-controlled feasibility study}.
\newblock \bibinfo{journal}{Korean journal of radiology} \bibinfo{volume}{19},
  \bibinfo{pages}{568--577}.
\bibitem[{Hariharan et~al.(2014)Hariharan, Arbel{\'a}ez, Girshick and
  Malik}]{hariharan2014simultaneous}
\bibinfo{author}{Hariharan, B.}, \bibinfo{author}{Arbel{\'a}ez, P.},
  \bibinfo{author}{Girshick, R.}, \bibinfo{author}{Malik, J.},
  \bibinfo{year}{2014}.
\newblock \bibinfo{title}{Simultaneous detection and segmentation}, in:
  \bibinfo{booktitle}{European Conference on Computer Vision},
  \bibinfo{organization}{Springer}. pp. \bibinfo{pages}{297--312}.
\bibitem[{He et~al.(2017)He, Gkioxari, Doll{\'a}r and Girshick}]{he2017mask}
\bibinfo{author}{He, K.}, \bibinfo{author}{Gkioxari, G.},
  \bibinfo{author}{Doll{\'a}r, P.}, \bibinfo{author}{Girshick, R.},
  \bibinfo{year}{2017}.
\newblock \bibinfo{title}{Mask r-cnn}, in: \bibinfo{booktitle}{Proceedings of
  the IEEE international conference on computer vision}, pp.
  \bibinfo{pages}{2961--2969}.
\bibitem[{Id{\'e}e et~al.(2006)Id{\'e}e, Port, Raynal, Schaefer, Le~Greneur and
  Corot}]{idee2006clinical}
\bibinfo{author}{Id{\'e}e, J.M.}, \bibinfo{author}{Port, M.},
  \bibinfo{author}{Raynal, I.}, \bibinfo{author}{Schaefer, M.},
  \bibinfo{author}{Le~Greneur, S.}, \bibinfo{author}{Corot, C.},
  \bibinfo{year}{2006}.
\newblock \bibinfo{title}{Clinical and biological consequences of
  transmetallation induced by contrast agents for magnetic resonance imaging: a
  review}.
\newblock \bibinfo{journal}{Fundamental \& clinical pharmacology}
  \bibinfo{volume}{20}, \bibinfo{pages}{563--576}.
\bibitem[{Kele and van~der Jagt(2010)}]{kele2010diffusion}
\bibinfo{author}{Kele, P.G.}, \bibinfo{author}{van~der Jagt, E.J.},
  \bibinfo{year}{2010}.
\newblock \bibinfo{title}{Diffusion weighted imaging in the liver}.
\newblock \bibinfo{journal}{World journal of gastroenterology: WJG}
  \bibinfo{volume}{16}, \bibinfo{pages}{1567}.
\bibitem[{Kierans et~al.(2016)Kierans, Kang and
  Rosenkrantz}]{kierans2016diagnostic}
\bibinfo{author}{Kierans, A.S.}, \bibinfo{author}{Kang, S.K.},
  \bibinfo{author}{Rosenkrantz, A.B.}, \bibinfo{year}{2016}.
\newblock \bibinfo{title}{The diagnostic performance of dynamic
  contrast-enhanced mr imaging for detection of small hepatocellular carcinoma
  measuring up to 2 cm: a meta-analysis}.
\newblock \bibinfo{journal}{Radiology} \bibinfo{volume}{278},
  \bibinfo{pages}{82--94}.
\bibitem[{Kim et~al.(2020a)Kim, Min, Kim, Shin and Lee}]{kim2020detection}
\bibinfo{author}{Kim, J.}, \bibinfo{author}{Min, J.H.}, \bibinfo{author}{Kim,
  S.K.}, \bibinfo{author}{Shin, S.Y.}, \bibinfo{author}{Lee, M.W.},
  \bibinfo{year}{2020}a.
\newblock \bibinfo{title}{Detection of hepatocellular carcinoma in
  contrast-enhanced magnetic resonance imaging using deep learning classifier:
  A multi-center retrospective study}.
\newblock \bibinfo{journal}{Scientific Reports} \bibinfo{volume}{10},
  \bibinfo{pages}{1--11}.
\bibitem[{Kim et~al.(2020b)Kim, Lee, Baek and Yun}]{kim2020diagnostic}
\bibinfo{author}{Kim, J.S.}, \bibinfo{author}{Lee, J.K.},
  \bibinfo{author}{Baek, S.Y.}, \bibinfo{author}{Yun, H.I.},
  \bibinfo{year}{2020}b.
\newblock \bibinfo{title}{Diagnostic performance of a minimized protocol of
  non-contrast mri for hepatocellular carcinoma surveillance}.
\newblock \bibinfo{journal}{Abdominal Radiology} \bibinfo{volume}{45},
  \bibinfo{pages}{211--219}.
\bibitem[{Kim et~al.(2012)Kim, Lee, Park, Kim, Rhim and
  Choi}]{kim2012hypovascular}
\bibinfo{author}{Kim, Y.K.}, \bibinfo{author}{Lee, W.J.},
  \bibinfo{author}{Park, M.J.}, \bibinfo{author}{Kim, S.H.},
  \bibinfo{author}{Rhim, H.}, \bibinfo{author}{Choi, D.}, \bibinfo{year}{2012}.
\newblock \bibinfo{title}{Hypovascular hypointense nodules on hepatobiliary
  phase gadoxetic acid--enhanced mr images in patients with cirrhosis:
  potential of dw imaging in predicting progression to hypervascular hcc}.
\newblock \bibinfo{journal}{Radiology} \bibinfo{volume}{265},
  \bibinfo{pages}{104--114}.
\bibitem[{Leng et~al.(2018)Leng, Liu, Zhang, Quan and Cui}]{leng2018context}
\bibinfo{author}{Leng, J.}, \bibinfo{author}{Liu, Y.}, \bibinfo{author}{Zhang,
  T.}, \bibinfo{author}{Quan, P.}, \bibinfo{author}{Cui, Z.},
  \bibinfo{year}{2018}.
\newblock \bibinfo{title}{Context-aware u-net for biomedical image
  segmentation}, in: \bibinfo{booktitle}{2018 IEEE International Conference on
  Bioinformatics and Biomedicine (BIBM)}, \bibinfo{organization}{IEEE}. pp.
  \bibinfo{pages}{2535--2538}.
\bibitem[{Marckmann et~al.(2006)Marckmann, Skov, Rossen, Dupont, Damholt, Heaf
  and Thomsen}]{marckmann2006nephrogenic}
\bibinfo{author}{Marckmann, P.}, \bibinfo{author}{Skov, L.},
  \bibinfo{author}{Rossen, K.}, \bibinfo{author}{Dupont, A.},
  \bibinfo{author}{Damholt, M.B.}, \bibinfo{author}{Heaf, J.G.},
  \bibinfo{author}{Thomsen, H.S.}, \bibinfo{year}{2006}.
\newblock \bibinfo{title}{Nephrogenic systemic fibrosis: suspected causative
  role of gadodiamide used for contrast-enhanced magnetic resonance imaging}.
\newblock \bibinfo{journal}{Journal of the American Society of Nephrology}
  \bibinfo{volume}{17}, \bibinfo{pages}{2359--2362}.
\bibitem[{Pang et~al.(2019)Pang, Su, Leung, Nachum, Chen, Feng and
  Li}]{pang2019direct}
\bibinfo{author}{Pang, S.}, \bibinfo{author}{Su, Z.}, \bibinfo{author}{Leung,
  S.}, \bibinfo{author}{Nachum, I.B.}, \bibinfo{author}{Chen, B.},
  \bibinfo{author}{Feng, Q.}, \bibinfo{author}{Li, S.}, \bibinfo{year}{2019}.
\newblock \bibinfo{title}{Direct automated quantitative measurement of spine by
  cascade amplifier regression network with manifold regularization}.
\newblock \bibinfo{journal}{Medical image analysis} \bibinfo{volume}{55},
  \bibinfo{pages}{103--115}.
\bibitem[{Piana et~al.(2011)Piana, Trinquart, Meskine, Barrau, Van~Beers and
  Vilgrain}]{piana2011new}
\bibinfo{author}{Piana, G.}, \bibinfo{author}{Trinquart, L.},
  \bibinfo{author}{Meskine, N.}, \bibinfo{author}{Barrau, V.},
  \bibinfo{author}{Van~Beers, B.}, \bibinfo{author}{Vilgrain, V.},
  \bibinfo{year}{2011}.
\newblock \bibinfo{title}{New mr imaging criteria with a diffusion-weighted
  sequence for the diagnosis of hepatocellular carcinoma in chronic liver
  diseases}.
\newblock \bibinfo{journal}{Journal of hepatology} \bibinfo{volume}{55},
  \bibinfo{pages}{126--132}.
\bibitem[{Ren et~al.(2015)Ren, He, Girshick and Sun}]{ren2015faster}
\bibinfo{author}{Ren, S.}, \bibinfo{author}{He, K.}, \bibinfo{author}{Girshick,
  R.}, \bibinfo{author}{Sun, J.}, \bibinfo{year}{2015}.
\newblock \bibinfo{title}{Faster r-cnn: Towards real-time object detection with
  region proposal networks}, in: \bibinfo{booktitle}{Advances in neural
  information processing systems}, pp. \bibinfo{pages}{91--99}.
\bibitem[{Ronneberger et~al.(2015)Ronneberger, Fischer and
  Brox}]{ronneberger2015u}
\bibinfo{author}{Ronneberger, O.}, \bibinfo{author}{Fischer, P.},
  \bibinfo{author}{Brox, T.}, \bibinfo{year}{2015}.
\newblock \bibinfo{title}{U-net: Convolutional networks for biomedical image
  segmentation}, in: \bibinfo{booktitle}{International Conference on Medical
  image computing and computer-assisted intervention},
  \bibinfo{organization}{Springer}. pp. \bibinfo{pages}{234--241}.
\bibitem[{Sobel and Feldman(1968)}]{sobel19683x3}
\bibinfo{author}{Sobel, I.}, \bibinfo{author}{Feldman, G.},
  \bibinfo{year}{1968}.
\newblock \bibinfo{title}{A 3x3 isotropic gradient operator for image
  processing}.
\newblock \bibinfo{journal}{a talk at the Stanford Artificial Project in} ,
  \bibinfo{pages}{271--272}.
\bibitem[{Van~Griethuysen et~al.(2017)Van~Griethuysen, Fedorov, Parmar, Hosny,
  Aucoin, Narayan, Beets-Tan, Fillion-Robin, Pieper and
  Aerts}]{van2017computational}
\bibinfo{author}{Van~Griethuysen, J.J.}, \bibinfo{author}{Fedorov, A.},
  \bibinfo{author}{Parmar, C.}, \bibinfo{author}{Hosny, A.},
  \bibinfo{author}{Aucoin, N.}, \bibinfo{author}{Narayan, V.},
  \bibinfo{author}{Beets-Tan, R.G.}, \bibinfo{author}{Fillion-Robin, J.C.},
  \bibinfo{author}{Pieper, S.}, \bibinfo{author}{Aerts, H.J.},
  \bibinfo{year}{2017}.
\newblock \bibinfo{title}{Computational radiomics system to decode the
  radiographic phenotype}.
\newblock \bibinfo{journal}{Cancer research} \bibinfo{volume}{77},
  \bibinfo{pages}{e104--e107}.
\bibitem[{Vandecaveye et~al.(2009)Vandecaveye, De~Keyzer, Verslype, De~Beeck,
  Komuta, Topal, Roebben, Bielen, Roskams, Nevens
  et~al.}]{vandecaveye2009diffusion}
\bibinfo{author}{Vandecaveye, V.}, \bibinfo{author}{De~Keyzer, F.},
  \bibinfo{author}{Verslype, C.}, \bibinfo{author}{De~Beeck, K.O.},
  \bibinfo{author}{Komuta, M.}, \bibinfo{author}{Topal, B.},
  \bibinfo{author}{Roebben, I.}, \bibinfo{author}{Bielen, D.},
  \bibinfo{author}{Roskams, T.}, \bibinfo{author}{Nevens, F.}, et~al.,
  \bibinfo{year}{2009}.
\newblock \bibinfo{title}{Diffusion-weighted mri provides additional value to
  conventional dynamic contrast-enhanced mri for detection of hepatocellular
  carcinoma}.
\newblock \bibinfo{journal}{European radiology} \bibinfo{volume}{19},
  \bibinfo{pages}{2456--2466}.
\bibitem[{Wu and Nevatia(2007)}]{wu2007simultaneous}
\bibinfo{author}{Wu, B.}, \bibinfo{author}{Nevatia, R.}, \bibinfo{year}{2007}.
\newblock \bibinfo{title}{Simultaneous object detection and segmentation by
  boosting local shape feature based classifier}, in: \bibinfo{booktitle}{2007
  IEEE Conference on Computer Vision and Pattern Recognition},
  \bibinfo{organization}{IEEE}. pp. \bibinfo{pages}{1--8}.
\bibitem[{Wu et~al.(2019)Wu, Liu, Cui, Chen, Song and Xie}]{wu2019radiomics}
\bibinfo{author}{Wu, J.}, \bibinfo{author}{Liu, A.}, \bibinfo{author}{Cui, J.},
  \bibinfo{author}{Chen, A.}, \bibinfo{author}{Song, Q.}, \bibinfo{author}{Xie,
  L.}, \bibinfo{year}{2019}.
\newblock \bibinfo{title}{Radiomics-based classification of hepatocellular
  carcinoma and hepatic haemangioma on precontrast magnetic resonance images}.
\newblock \bibinfo{journal}{BMC medical imaging} \bibinfo{volume}{19},
  \bibinfo{pages}{23}.
\bibitem[{Xiao et~al.(2019)Xiao, Zhao, Qiang, Chong, Yang, Kazihise, Chen and
  Li}]{xiao2019radiomics}
\bibinfo{author}{Xiao, X.}, \bibinfo{author}{Zhao, J.}, \bibinfo{author}{Qiang,
  Y.}, \bibinfo{author}{Chong, J.}, \bibinfo{author}{Yang, X.},
  \bibinfo{author}{Kazihise, N.G.F.}, \bibinfo{author}{Chen, B.},
  \bibinfo{author}{Li, S.}, \bibinfo{year}{2019}.
\newblock \bibinfo{title}{Radiomics-guided gan for segmentation of liver tumor
  without contrast agents}, in: \bibinfo{booktitle}{International Conference on
  Medical Image Computing and Computer-Assisted Intervention},
  \bibinfo{organization}{Springer}. pp. \bibinfo{pages}{237--245}.
\bibitem[{Xu et~al.(2009)Xu, Yan, Wang, Lin and Ji}]{xu2009added}
\bibinfo{author}{Xu, P.J.}, \bibinfo{author}{Yan, F.H.}, \bibinfo{author}{Wang,
  J.H.}, \bibinfo{author}{Lin, J.}, \bibinfo{author}{Ji, Y.},
  \bibinfo{year}{2009}.
\newblock \bibinfo{title}{Added value of breathhold diffusion-weighted mri in
  detection of small hepatocellular carcinoma lesions compared with dynamic
  contrast-enhanced mri alone using receiver operating characteristic curve
  analysis}.
\newblock \bibinfo{journal}{Journal of Magnetic Resonance Imaging: An Official
  Journal of the International Society for Magnetic Resonance in Medicine}
  \bibinfo{volume}{29}, \bibinfo{pages}{341--349}.
\bibitem[{Yu et~al.(1999)Yu, Kim, Kim, Lee and Yoo}]{yu1999contrast}
\bibinfo{author}{Yu, J.S.}, \bibinfo{author}{Kim, K.W.}, \bibinfo{author}{Kim,
  E.K.}, \bibinfo{author}{Lee, J.T.}, \bibinfo{author}{Yoo, H.S.},
  \bibinfo{year}{1999}.
\newblock \bibinfo{title}{Contrast enhancement of small hepatocellular
  carcinoma: usefulness of three successive early image acquisitions during
  multiphase dynamic mr imaging.}
\newblock \bibinfo{journal}{AJR. American journal of roentgenology}
  \bibinfo{volume}{173}, \bibinfo{pages}{597--604}.
\bibitem[{Yushkevich et~al.(2006)Yushkevich, Piven, Hazlett, Smith, Ho, Gee and
  Gerig}]{yushkevich2006user}
\bibinfo{author}{Yushkevich, P.A.}, \bibinfo{author}{Piven, J.},
  \bibinfo{author}{Hazlett, H.C.}, \bibinfo{author}{Smith, R.G.},
  \bibinfo{author}{Ho, S.}, \bibinfo{author}{Gee, J.C.},
  \bibinfo{author}{Gerig, G.}, \bibinfo{year}{2006}.
\newblock \bibinfo{title}{User-guided 3d active contour segmentation of
  anatomical structures: significantly improved efficiency and reliability}.
\newblock \bibinfo{journal}{Neuroimage} \bibinfo{volume}{31},
  \bibinfo{pages}{1116--1128}.
\bibitem[{Zhao et~al.(2020)Zhao, Li, Kassam, Howey, Chong, Chen and
  Li}]{zhao2020tripartite}
\bibinfo{author}{Zhao, J.}, \bibinfo{author}{Li, D.}, \bibinfo{author}{Kassam,
  Z.}, \bibinfo{author}{Howey, J.}, \bibinfo{author}{Chong, J.},
  \bibinfo{author}{Chen, B.}, \bibinfo{author}{Li, S.}, \bibinfo{year}{2020}.
\newblock \bibinfo{title}{Tripartite-gan: Synthesizing liver contrast-enhanced
  mri to improve tumor detection}.
\newblock \bibinfo{journal}{Medical Image Analysis} , \bibinfo{pages}{101667}.

\end{thebibliography}
\end{document}